# Fabrication of angstrom-scale two-dimensional channels for mass transport

Ankit Bhardwaj[1,2], Marcos Vinicius Surmani Martins[1,2], Yi You[1,2], Ravalika Sajja[1,2], Max Rimmer[1,2], Solleti Goutham[1,2], Rongrong Qi[1,2], Sidra Abbas Dar[1,2], Boya Radha[1,2*] and Ashok Keerthi[1,3*]

[1] National Graphene Institute, The University of Manchester, Manchester M13 9PL, United Kingdom.

[2] Department of Physics and Astronomy, School of Natural Sciences, The University of Manchester, Manchester M13 9PL, United Kingdom.

[3] Department of Chemistry, School of Natural Sciences, The University of Manchester, Manchester M13 9PL, United Kingdom.

*Please address all correspondence to: Ashok Keerthi (ashok.keerthi@manchester.ac.uk) or Radha Boya (radha.boya@manchester.ac.uk)

**Abstract**

Fluidic channels at atomic scales regulate cellular trafficking and molecular filtration across membranes and thus play crucial roles in the functioning of living systems. However, constructing synthetic channels experimentally at these scales has been a significant challenge due to the limitations in nanofabrication techniques and the surface roughness of the commonly used materials. Angstrom (Å)-scale slit-like channels address this challenge, as these can be made with precise control over their dimensions and can be used to study the fluidic properties of gases, ions and water at unprecedented scales. Here, we provide a detailed fabrication method of the two-dimensional (2D) Å-scale channels, which can be assembled as a single channel or up to hundreds of channels made with atomic scale precision using layered crystals. The procedure includes the fabrication of the substrate, flake, spacer layer, flake transfers, van der Waals assembly, and post-processing. We further explain how to perform molecular transport measurements with the Å-channels, for the development of methods directed at unravelling interesting and anomalous phenomena that help shed light on the physics of nanofluidic transport systems. The procedure requires a total of 1 – 2 weeks for the fabrication of the 2D channel device and is suitable for users with prior experience in clean room working environments and nanofabrication.

**Introduction**

Microfluidics, the study of fluid transport at microscale, has evolved as a major research field over the last three decades with important applications in diagnostics, optical imaging, and the local cooling of semiconductor chips, amongst others [1-3]. Particularly, after developments in microfabrication techniques used in microelectronics, microfluidic devices have gained momentum and expanded to make contributions to biotechnology, flow chemistry, membrane science and energy storage applications [1,4]. Nanoelectronics also emerged as a subfield due to the increased interest in device miniaturization. Advances in the fabrication processes allowed to push the boundaries of microfluidics to nanofluidics [2,4]. Nanofluidics is not simply about downscaling the dimensions already found in microfluidic systems, but rather develops know-how about the nanoscale effects of molecular interactions. Reducing the dimensions further to the atomic scale, i.e., roughly a few angstroms, opens avenues to atomic-scale confined fluidics, which can be called *angstrofluidics*. The method to fabricate two-dimensional (2D) capillaries with Å-scale precision is of wide interest, as these devices mark an important milestone towards the development of angstrofluidics [2,4].

In recent decades, the field of nanofluidics has seen notable applications in membrane science, surface science and biosciences [2], as an inspiration is derived from biological systems (e.g., the human body and plants). Within the natural systems, the fluid flow through nanoconfined space is normally governed by a series of ordered and manifold functionalities of nanofluidic channels [5,6]. For instance, the aquaporin channel with a pore size close to the size of water molecule can only accommodate a single file water transport [5]. To understand the mechanisms found in nature, relentless efforts have been made to build platforms mimicking these nanofluidic transport processes. There are numerous materials that hold promise, like zero dimensional (0D) nanoporous materials (e.g., metal organic frameworks [7-9] and nanoporous graphene [10-12]), one-dimensional (1D) materials such as carbon nanotubes [13-15] and boron nitride nanotubes [16,17]), and two-dimensional (2D) laminate materials, for example graphene oxide membranes [18-20] and MXene membranes [21-23]). Owing to the emerging nanofluidic systems, many intriguing physical phenomena have been explored like the ultrafast water flow along carbon nanotube (CNT) channel [24] and the observation of ionic coulomb blockade [25]. However, experimental demonstrations of the prominent behaviors for nanofluids are realized by very few platforms, whereas most of the existing platforms confront technical difficulties in achieving precise confinement control. Therefore, a robust method to fabricate a device with Å-scale precision is crucial for the comprehensive understanding of nanofluidics.

**Methods of making nanostructures**

There are two general routes for the fabrication of nanostructures: top-down and bottom-up approaches [26,27]. Both approaches can be used to create nanopores or nanoporous materials that are suitable for investigating nanofluidic properties.

Top-down approaches involve breaking, reducing or disintegrating the bulk material in a controlled manner to obtain the desired nanostructures. This approach uses drilling techniques such as electron beam lithography (EBL), reactive ion etching (RIE), focussed ion beam (FIB) etc., to remove material from bulk samples. These techniques can be applied to create pores in bulk non-porous materials, with tunable size and shapes [28]. For instance, solid state nanopores that are mechanically robust and stable [28] have been produced on materials like silicon nitride ($SiN_x$) [29-32], silicon oxide ($SiO_2$) [33], silicon carbide [34], and in 2D materials like suspended graphene [35,36], hexagonal boron nitride (hBN) [37,38], tungsten disulfide [39] and MXene [40]. Additionally, controlled wet etching with aqueous solutions of potassium hydroxide (KOH) can be used to create nanopores in Si at a wafer scale [28].

Bottom-up approaches involve bringing atoms, molecules or fragments together to assemble the nanostructures. These methods can be used to assemble nanoporous and nanofluidic devices from smaller components. Assembling 2D porous polymers by using reactions that occur only at the interface between two substances [41-45], and the growth of metal organic frameworks from their constituent elements [46] are some of the examples of bottom-up approaches. Other bottom-up methods include chemical vapour deposition (CVD) for creating vacancy defects in thin-films as intrinsic pores, and vacuum assisted filtration to form nanoporous membranes and 2D laminates from dispersed solutions of graphene oxide, vermiculite and MXenes [19,47-50].

An important method of fabrication specific to devices made from 2D materials, such as graphene and hBN, is the process of stacking individual layers into van der Waals heterostructures [51]. Such stacks of individual layers with different 2D crystals are held together by van der Waals forces, hence named as van der Waals heterostructures. These stacks are generally manufactured by picking up one flake (flake could be one or few layers of 2D crystal) with assistance of a polymer layer, placing this flake in the desired location on top of another flake and subsequently dissolving the polymer [52,53]. Repeating this step with multiple flakes of different materials can produce highly configurable devices in a process that has been described as "atomic-scale Lego" [51]. Devices of this nature have been fabricated to study magnetic [54], electronic [55] and optical properties [56]. This protocol describes the process of producing van der Waals heterostructures with Å-scale fluidic channels (see Fig. 1) capable of being used for molecular transport, ionic current measurements, and investigation of fundamental properties of fluidic matter under atomic scale confinement.

## Angstrom (Å)-scale channels

Our Å-scale fluidic channel devices are made from tri-crystal stacks of 2D materials where the middle 2D crystal layer is patterned by electron beam lithography to make array of parallel stripes or spacers. This forms a gap, i.e., an Å-scale 2D channel between the top and bottom crystal where fluids can flow through as depicted in Fig. 1. The spacer flake can be single layer to few layers thick 2D material, which defines the height of the capillary. These Å-channels devices not only provide atomic scale tunability of dimensions, but also ensure atomically smooth channel walls. These tri-crystal stacks are transferred onto a micrometre sized, rectangular hole in a free-standing silicon nitride ($SiN_x$) membrane on silicon (Si) substrate to fabricate 2D channel devices (see Fig. 2 for complete fabrication flow chart). They are meticulously stacked such that fluids can flow from one side of the membrane to the other side only through the capillaries (Fig. 3). We fabricate devices with single and multiple channels and variable channel dimensions where channel's height ($h$) ≈ 0.34 nm – 300 nm, width ($w$) ≈ 50 nm - 150 nm and length $L$ ≈ 0.5 µm – 15 µm by using the techniques in this protocol with high reproducibility and flexibility in design. With high precision in dimensions, nanofluidic devices have been demonstrated to possess diverse new properties, including fast water flow and ballistic gas transport, non-linear electrokinetic transport, and high proton mobility under confinement [57-61]. They also display anomalies such as lower dielectric constant of confined water compared to bulk [62]. In addition to small molecules and ions, a double-stranded DNA, a biopolymer with few thousands of base pairs, is able to translocate through these 2D channels with uniform velocity without noticeable frictional interactions of DNA with the confining graphene surfaces [63].

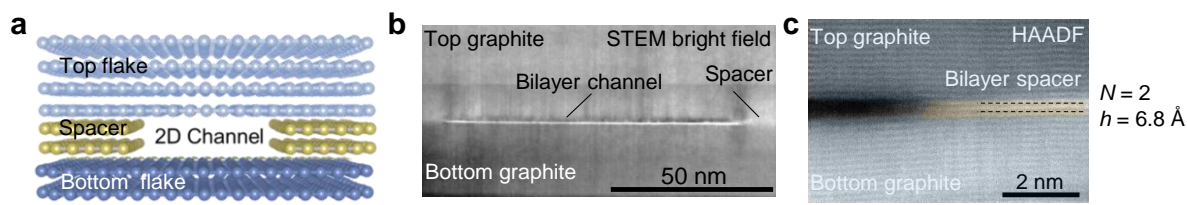

**Fig. 1| Angstrom-scale channels. a,** Schematic illustration of an Å-channel. It consists of three components: a top flake, a bottom flake and a spacer, which forms the walls of a channel that is two atoms thick. **b,** Cross-sectional bright field image of a bilayer channel (height ≈ 6.8 Å) in a scanning transmission electron microscope. This channel is made from top and bottom crystals of graphite and spacers of bilayer graphene. **c,** High-angle annular dark field (HAADF) image of the edge of a bilayer channel in panel **b** [57].

## Overview of the procedure

The procedure for making Å-channel device is divided into five parts: substrate fabrication (Steps 1-22), flake preparation (Steps 23-30), spacer layer fabrication (Step 31-49), flake transfers and van der Waals assembly (Steps 50-80), and post-processing of tri-crystal stack and gold patch deposition (Steps 81-107). A brief fabrication flow chart is shown in Fig. 2 and graphical overview of steps is shown in Fig. 3. For efficient production of devices, fabrication of substrates and preparation of flakes are done at wafer scale, whereas material transfer, etching and post treatment are processed at a chip scale (typically 2 cm × 2 cm).

Fabrication of such 2D channels can also be made on any substrate by simply stacking bottom, spacer, and top layers of any 2D crystals where entry and exit of channels are on the same plane. However, in order to measure flow through these capillaries, connections must be made directly to the channel openings, which is challenging in terms of fabrication and alignment. Our Å-channels, present on a $SiN_x$/Si wafer chip with out-of-plane entry/exit, allow mass transport measurements using detection/characterization techniques (e.g., mass spectrometer, electrical meters) built on either side of the two chambers and separated by channels.

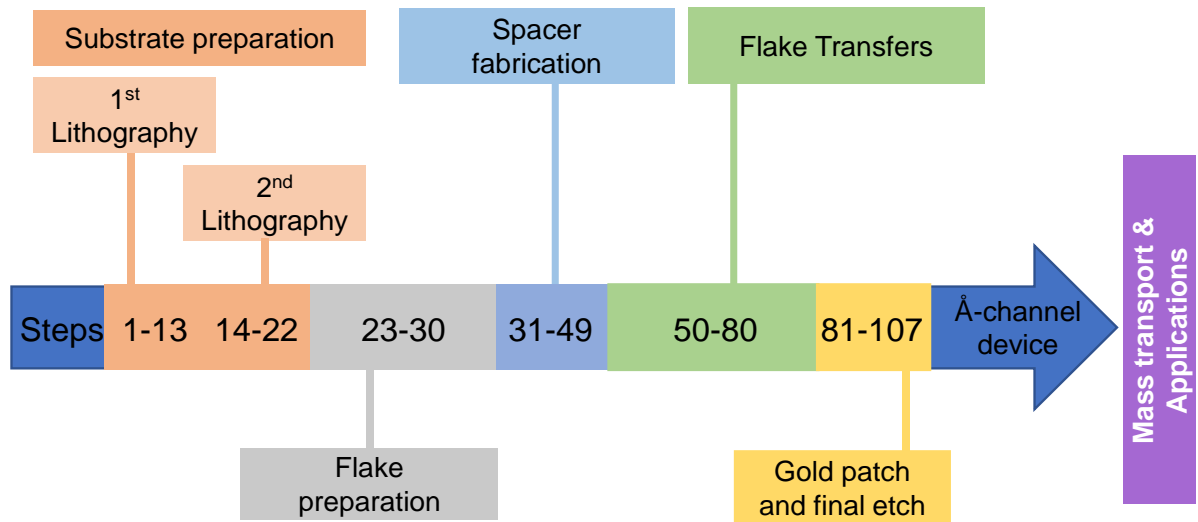

**Fig. 2| Flow chart of the Å-scale device fabrication.** The whole fabrication procedure includes the following: substrate preparation (Steps 1-22), flake preparation (Steps 23-30), spacer fabrication (Step 31-49), transfer of bottom, spacers, and top flake (Steps 50-80) and patterning gold patch and final etch of exposed 2D crystals (Steps 81-107).

The Å-channel device is made of a tri-crystal stack on a $SiN_x$ membrane with a hole supported by a Si substrate (Fig. 3a). The fabrication starts with preparation of a freestanding $SiN_x$ membrane of approximately 150 µm × 150 µm in size using commercially available Si wafers with 500 nm thick $SiN_x$ on both sides (Steps 1-13). A rectangular hole (~3 µm × ~25 µm) is made in the free-standing membrane using photolithography and reactive ion etching (RIE) respectively (Steps 14-22). Then a graphite crystal (or any other 2D crystal as bottom layer) with thickness > 10 nm is deposited to seal the opening using the transfer method (described in section "Transfer of flakes").

Commercially available Si wafer with 290 nm thick $SiO_2$ on the one side polished Si substrate are used to exfoliate multilayer graphene flakes using micromechanical exfoliation, and flakes with desired thickness are selected to serve as a spacer layer (Steps 23-30). The spacer flake is patterned by EBL and oxygen plasma etching to create an array of parallel stripes of ~130 nm in width (Step 31-49). These spacer stripes are separated by few tens of nanometers to few microns so that they can be made using typical nanofabrication facilities with lithography and etching facilities. However, width of the channel and thickness of the top flake need to be optimized to prevent the collapse of the top layer into the channels. The spacer stripes are then transferred onto the bottom 2D crystal such that they are aligned perpendicular to the long side of the rectangular opening. Dry plasma etching is employed to drill through the bottom-spacer stack along the microhole, with the $SiN_x$ as a mask. Finally, another 2D crystal (70 nm to 150 nm thick) is transferred to serve as the top layer. After each transfer, the stack of 2D crystal layers is annealed at 350 °C to 400 °C for 4 hours to remove possible contamination (Step 81). The scanning electron microscope (SEM) images of a representative multichannel device are given in Fig. 3d (top view) with a defined channel length of $L$, width of $w$ and channel height, $h$. The cavity formed by the van der Waal tri-crystals stack is the so-called Å-channels where the channel height is solely controlled by the spacers' thickness (Fig. 1). This completes a set of 2D capillaries, such that their entries and exits are accessible from the opposite sides of the $SiN_x$/Si wafer (Fig. 3b, e) and this out-of-plane entry/exit is a key factor enabling mass transport measurements. To prevent the clogging or any blockage due to thin edges on the top crystal, a gold mask in rectangular patch pattern (gold patch) is deposited and final etching of excess top layer is done (Steps 81-107). This will also help to precisely define the channel length *L* (Fig 3c, f).

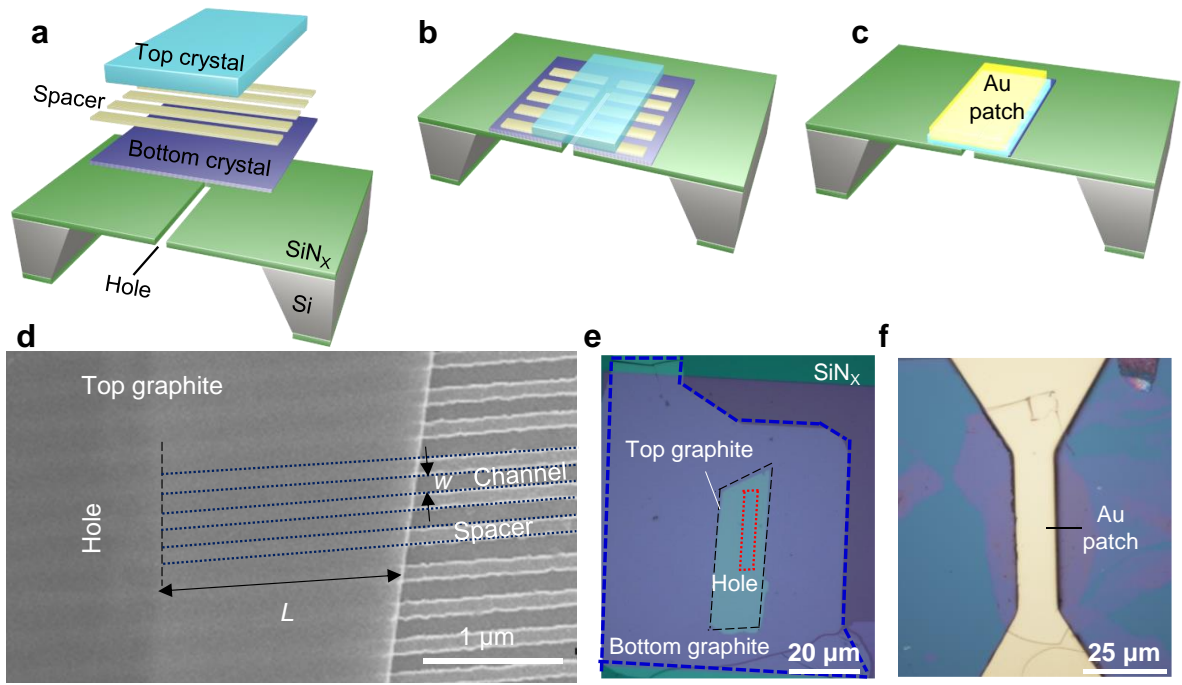

**Fig. 3| Overview of the fabrication process. a** and **b,** Schematic illustrations of Å-channel devices with a top, spacer and bottom on a hole in the $SiN_x$ membrane with a pre-etched hole. Firstly, the bottom crystal is transferred onto a $SiN_x$ hole followed by transfer of an array of spacers. The hole is extended up through the bottom-spacer stack by dry etching. Finally, the top crystal is transferred to cover the resulting aperture. **c,** Schematic of the final Å-channel device after gold patch. **d,** Top view of scanning electron microscopy image of a device. The spacers can be seen clearly (three of them are indicated by dotted lines and edge of the hole by dashed lines) where they are not covered by the top graphite crystal. **e,** Optical image of a typical Å-channel device with graphite walls on a $SiN_x$ membrane with hole. Top graphite is outlined with black dash line, bottom graphite is outlined with blue dotted line and hole is outlined with red colour round dots. **f,** Optical image of an Å-channel device (different from **e**) after the gold patch deposition and lift-off process. (Image in **d** adapted from ref. [57]).

## Applications

Å-channel devices can be used to understand fluidics at the smallest scale possible. Here, we will give details of three types of flow measurements that use the Å-scale channels to analyze important transport properties.

### *Ballistic gas flows*

Measurements from Å-channel devices suggest that ballistic transport can occur for gas molecules flowing through capillaries, leading to higher than predicted gas flow rates. In extremely confined conduits, gas molecules mean free path is much larger than the confining dimension, resulting in a free molecular flow regime, where gas molecules collide more with the walls than each other. The dynamics of gas flow in free molecular regime is described using Knudsen diffusion theory [64-67]. Membranes with molecular-scale dimensions typically have permeation rate of the gases in agreement with the flow predictions of the Knudsen theory [11,68,69]. However, there are exceptions to Knudsen diffusion theory.

In devices that reported enhanced gas flux compared to Knudsen estimates, this enhancement was attributed to the effect of specular (mirror-like) scattering [70,71] or partially specular reflection [72,73]. Gas flow through Å-channel devices has been shown to be several orders of magnitude higher than that predicted by Knudsen theory. The Å-channels have atomic-scale confinement and atomically smooth walls, which facilitate free molecular flow. The enhanced gas flow in 2D capillaries is attributed to near-specular reflection (Fig. 4a), which leads to ballistic transport and frictionless gas flow [59].

*Ultrafast water permeation*

We can use the Å-channel devices to investigate water permeation under confinement. Understanding the water transport mechanisms in a nanocapillary is essential for better design of devices for water desalination, purification and building biomimetic system. Materials like atomically thin nanoporous membranes and carbon nanotubes are promising platforms with useful properties and a high potential for future applications. Theoretical simulations have demonstrated that nanoporous graphene and $MoS_2$, with pores terminated by hydrophilic groups could reduce the energy barrier, enabling enhanced water transport [74-77]. Another example is carbon nanotubes which show rapid water flow, attributed to the large slip length (3 μm to 70 μm) and a low-friction to water on its hydrophobic surface [24,78-80]. Though enormous advances have been made, most studies are only limited to the simulations, lacking experimental observations. The channels made using 2D materials as described in this protocol represents a prototype nanofluidic system to deepen the understanding of nanoconfined water transport. The schematic of water flow in the 2D capillary device at 100% relative humidity level is illustrated in Fig. 4b. When water enters the nanocapillary, it condenses due to the capillary condensation effect [57,81,82] and flows as liquid. Analogous to CNTs, the weak water-graphitic wall interaction in our capillary device affirms a slip length of ~ 60 nm to 100 nm, exhibiting a frictionless surface for water flow [57]. In addition to the large slip length, the high capillary pressure is another factor, which facilitates water transport. As the meniscus extends outside the channels, this capillary pressure continuously pulls the liquid water through the channel at an ultrafast speed of 1 m/s [57].

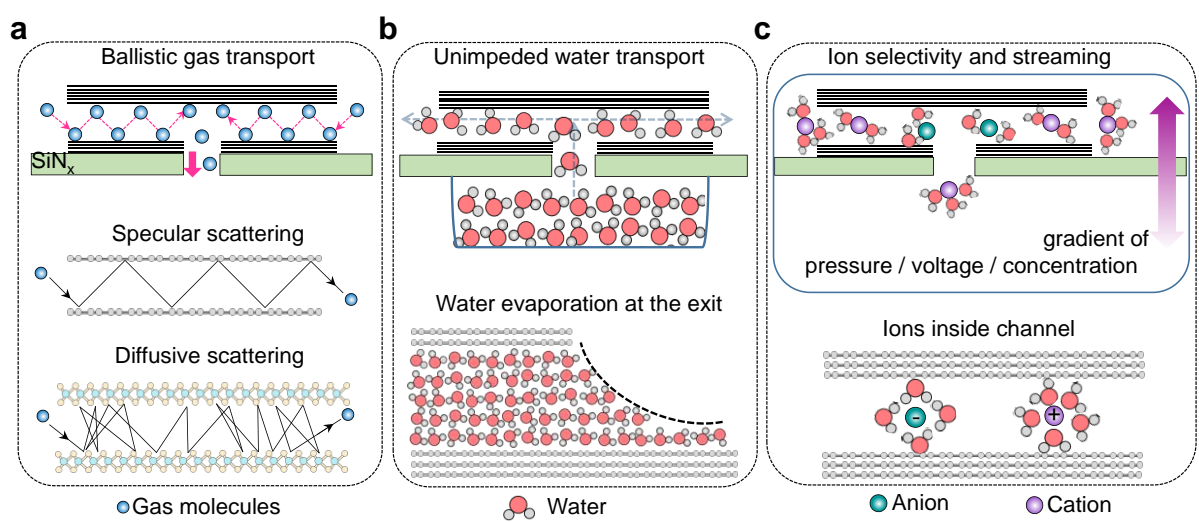

**Fig. 4| Molecular transport through Å-channel devices. a,** Schematics of ballistic gas transport through Å-channels. Ballistic transport is observed for devices with graphite and hBN walls. Specular scattering occurs when the channel's surface is atomically smooth such e.g., graphene, hBN walls, while diffusive scattering occurs when it is rougher such as atomic corrugations on $MoS_2$ walls. Shorter path lengths with faster flow rates are observed in the former case. **b,** Schematics of ultrafast water flow in Å-channels. The extended meniscus is outlined in the drawing, showing a thin film of water propagating along the graphite surface. The entire process involves transport of water vapour, liquid flow through the Å-channels, diffusion, and evaporation into air. **c,** Schematic of selective ion transport through the 2D confinement and asymmetry effects in hydration shell of cation and anion. Purple colour arrow

represents applied pressure, voltage, or concentration gradient across the channels (top) and the bottom schematic illustrates the squashed hydration shell around cations and anions.

*Steric ion exclusion and voltage gating of streaming currents*

The fabricated Å-channel devices are utilized for proton transport [60], size selective ion transport [58], producing pressure-induced ionic streaming currents, and also in voltage gating of the ion flows [61] as illustrated in Fig. 4c. Only protons ($H^+$) could flow in Å-channels fabricated with monolayer graphene spacers which are 3.4 Å thin, while small ions such as $Na^+$, $K^+$, $Li^+$, $Cl^-$ etc., are excluded [60]. When we use a two-layer spacer, it produces channels with height, ~ 6.8 Å, which allow permeation of the aforementioned ions with little resistance [58]. However, for those ions with hydrated diameter larger than the channel height, their mobility is suppressed in the 2D confinement proportionally to their hydrated diameters. This can be attributed to the squeezing of the large ions hydration shell as they flow through the Å-channels smaller than their hydrated diameter and are forced to undergo partial dehydration. Moreover, the arrangement of water molecules around the ions and their interaction with the channel walls can be a key factor to differentiate cation and anion in such Å-channels. Mechano-sensitive ion transport is also observed in Å-channels. By applying a pressure difference, these channels produce considerable amount of ionic current when water molecules along with ions are mechanically pushed through them [61]. Due to the selectivity of the Å-channels to one ion over the other in a salt (e.g., $K^+$ is more mobile than $Cl^-$ in KCl), it leads to a net current and potential across the channels, and this pressure-induced currents are known as streaming currents. Applying a voltage in addition to the pressure, leads to a transistor like effect where the streaming currents can be modulated several times by a fraction of voltage. This voltage gating is also sensitive to materials of the channel walls. For instance, with graphite walls, the streaming currents can be increased in a non-linear fashion, upto 20 times by 100 mV of voltage, whereas with hBN walls, the gating increases the streaming currents only by 2 to 3 times.

**Comparison with other methods and advantages**

The Å-channel device fabricated by using this protocol has many advantages over other methods. Firstly, we can produce capillaries made from 2D materials with dimensions which are precisely controlled, with channel heights ranging from 3.4 Å (monolayer graphene spacers) to few tens of nanometers, and height varied in steps of 3.4 Å or 6.6 Å depending on the spacer used (graphene or $MoS_2$). The presented Å-channel devices can be tailored with different wall materials from hydrophilic to hydrophobic and insulating to conducting or any combination of these properties, which is not possible through other existing nanofabrication methods. This protocol enables reproducible fabrication methods of Å-channels with control on channel dimensions as well as choice of wall material.

In comparison with nanofluidic devices such as nanotubes, lithographically patterned/etched channels on bulk materials, or 2D laminate membranes made from liquid phase exfoliated 2D materials (graphene oxide, $MoS_2$, MXene, etc.), our channels have atomically flat, pristine and mechanically robust channel walls. Our method offers the possibility to program the length, width, height, and number of channels of a device, starting with a single channel up to several hundreds.

Another, important advantage of this method is self-cleansing property of van der Waals assembly. When one 2D crystal is stacked over another by van der Waals interactions, contamination such as hydrocarbons are squeezed out [83], allowing cleaner interfaces and producing atomically flat channels. In the case of the thinnest channels, because of their small size, it is not entropically favourable for the hydrocarbons or polymers to enter or remain inside the channel.

**Limitations**

Fabrication of Å-channel devices using a 2D material spacer is a time intensive process and may result in medium-to-low yield (within 30% and 50%) of working devices. While membrane fabrication can be performed at wafer scale, all other steps must be performed on individual devices. Under ambient conditions, the channels may be contaminated due to adsorption of hydrocarbons on the surfaces of 2D materials, resulting in clogging of the channels over time. Hydrocarbons adsorption on most surfaces including 2D materials is unavoidable as proven by transmission electron microscopy studies of graphene surface in ultrahigh vacuum. Annealing at higher temperatures (350 °C to 400 °C) can reverse this blockage to some extent but not completely depending on the level of contamination. Other possible contamination can come from fabrication processes due to resist materials and process chemicals, which can introduce unwanted polymeric residue into the channels. This fabrication method may result in blocked 2D capillaries due to sagging of the top crystal layer into nanochannels. This can be addressed by choosing correct thickness of top 2D crystal as well as appropriate channel width. For example, the channel width of w ≈ 130 nm requires a top layer with more than at least 70 nm thick to prevent sagging into the channels which are > 1 nm height, and even thicker (100 nm to 150 nm) tops for < 1 nm thin channels.

Another limitation in the use of Å-channel devices made from 2D materials is that they cannot withstand a high voltage. With the application of high voltage (> 2 V), top crystal may delaminate from the rest of the stack which leads to heavy leakage in ionic currents or huge flux in molecular transport and hence destruction of device.

**Device fabrication and characterization**

*Substrate fabrication (Steps 1-22)*

The mechanical and chemical stability of the devices are critically dependent on the substrate and 2D materials used to fabricate the Å-channels. Hence, proper fabrication and device design is required for functioning devices with open channels that remain stable for several measurements. Typically, silicon-based substrates are used for on-chip fluidic devices due to their established usage in microfabrication, device integration and ease of use. We use $SiN_x$/Si wafer as substrate to make membranes which host Å-channels. A silicon nitride coated wafer is an ideal substrate for the membrane as Si has high etch selectivity compared to $SiN_x$ in KOH etching. In addition, $SiN_x$ substrate has acceptable surface roughness and reasonably good adhesion to the 2D materials.

Commercially available Si wafers with 500 nm thick $SiN_x$ on both sides (Fig. 5a), are used to make 500 nm thick free standing $SiN_x$ membranes. These $SiN_x$ membranes exhibit sufficient mechanical stability when blow-dried, while testing for gas flow measurements, or during the membrane transfer [84,85]. Here, a double side polished $SiN_x$/Si wafer is first cleaned with isopropyl alcohol (IPA) and acetone and dried under nitrogen flow. We have also used 100 nm thick $SiN_x$/Si wafers for some experiments, but 100 nm thick $SiN_x$ membranes are not as robust as 500 nm thick membranes in handling mechanical stress [86]. The wafer is coated with a positive photoresist S1813 (Fig. 5b) in a spin coater and soft baked at 110 °C (Table 1, Recipe R1). The patterns are made using photolithography by exposing to 405 nm wavelength light, then developed with the MF-319 developer. After this, the $SiN_x$ on the exposed regions of the wafer is dry etched using reactive ion etching (RIE) to form the windows (Fig. 5c). The wafer is cleaned in hot acetone (50 °C) and IPA to remove the resist layer (Fig. 5d) and then the exposed Si is etched by 30% KOH aqueous solution (Fig. 5e), thus leaving a suspended $SiN_x$ membrane. On the membrane, a microhole of the desired size is made via a second lithography step (Fig. 5f).

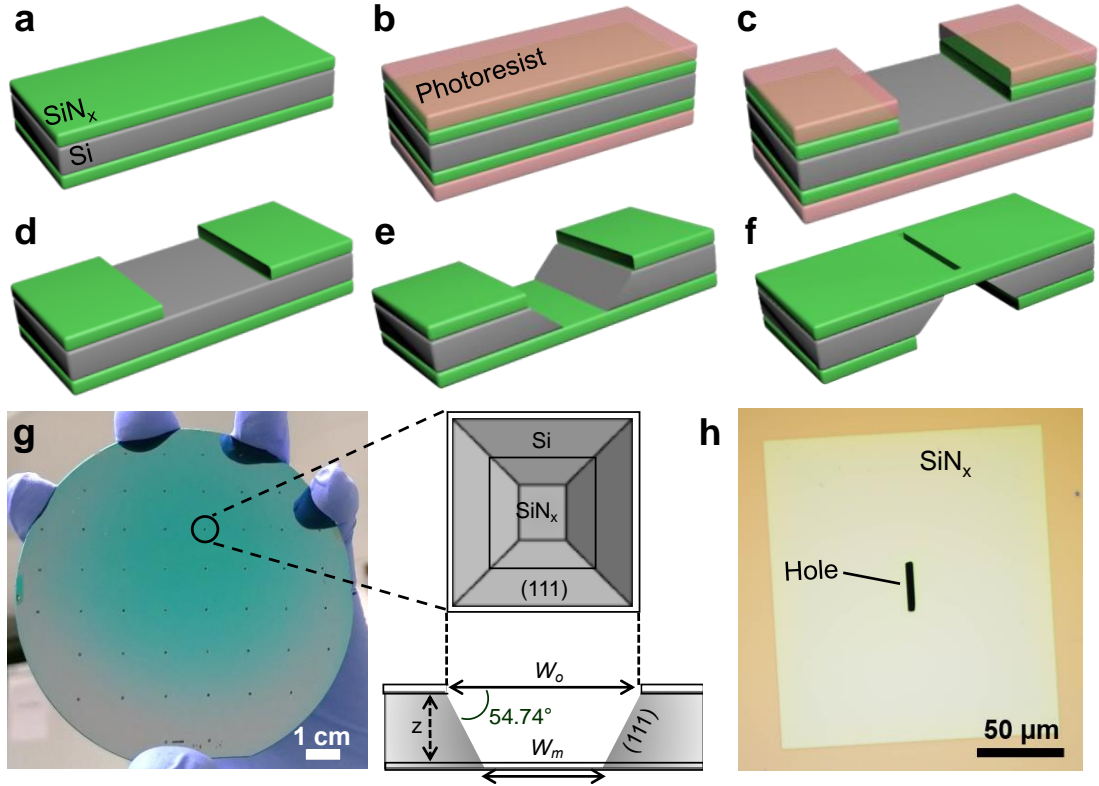

**Fig. 5| Schematic for the fabrication process of micrometre size SiN$_x$ holes at wafer scale. a,** Begin with a commercially available Si wafer with double side polished film of SiN$_x$ (thickness, 500 nm). **b,** Photoresist is coated on both sides of the wafer. **c,** The resist is patterned to form square windows *via* photolithography, then the resist is developed followed by dry etching through SiN$_x$. **d,** The resist is removed from the wafer by dissolving in hot acetone (50 °C) and washed with IPA. **e,** The exposed Si is etched in hot KOH (80 °C) which preferentially etches other planes compared to the <111> crystal planes, resulting in a free-standing SiN$_x$ membrane on the reverse side. **f,** A second round of photolithography and dry etching is used to produce micrometre sizes hole on the free-standing SiN$_x$ membrane. **g,** Digital photograph of wafer patterned with SiN$_x$ membranes, with a schematic of a membrane. $W_m$, $W_o$ and z are the desired width of the membrane, the width of initial aperture and the thickness of Si wafer, respectively. **h,** Optical image of the micro-hole on SiN$_x$ membrane.

Since KOH etches Si anisotropically with <111> direction being protected from etching, we can tailor dimensions of SiN$_x$ freestanding membrane [87] by following this equation [88]: 2·cot (54.74°) ≈ √2 and $W_o = W_m - √2$. The angle 54.74° is measured between horizontal level and <111> crystal plane etching direction [89]. The photoresist is coated on both sides of the wafer to keep SiN$_x$ surface as pristine as possible during dry etching. The SiN$_x$/Si wafer was patterned with the square/rectangular windows using laser writer, directly written on positive photoresist. The design of the windows was made using Clewin software. Typically, we design 800 μm × 800 μm windows on SiN$_x$/Si wafer to get 100 μm × 100 μm SiN$_x$ freestanding membranes using photolithography and dry etching as shown in Fig. 5g. Second photolithography is performed to make 3 μm × 25 μm hole in the middle of the membrane. As prepared membrane is then coated with S1813 photoresist and 3 μm × 25 μm hole is written on the SiN$_x$ membrane using laser writer, developed using MF-319 and SiN$_x$ is then etched in RIE. The photoresist is then cleaned by soaking the substrate in acetone, and IPA. The optical image of hole made on SiN$_x$ membrane is shown in Fig. 5h.

*Flake searching for bottom, spacer and top layer (Steps 23-30)*

The scotch tape (mechanical exfoliation) technique was first reported by Geim, Novoselov and co-workers to isolate graphene [90]. This is a simple technique to produce high quality flakes of 2D materials (graphene, or any other 2D materials like hBN, $MoS_2$, mica, etc.) with few defects. The mechanical exfoliation method is based on repeated peeling of the bulk 2D material (e.g., graphite) on the surface of tape and applying force on the crystal. The 2D material becomes thinner and thinner and exposes increasingly fewer layers or monolayer of the 2D material in the process.

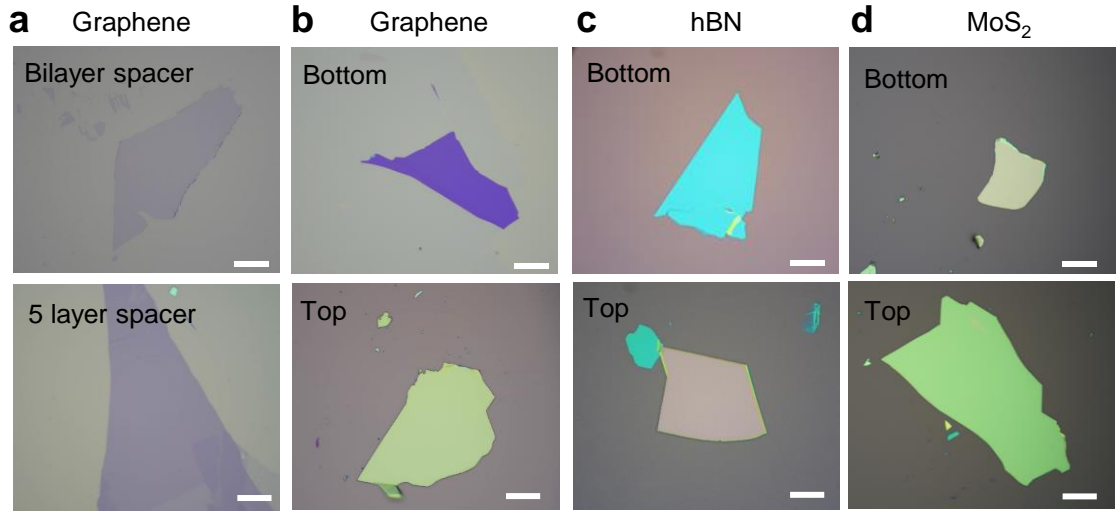

**Fig. 6|** Representative optical images of exfoliated flakes for preparing channels. **a**, bi-layer and 5-layer graphene which can be used as a spacer. **b**, graphene. **c**, hBN and **d**, $MoS_2$ flakes suitable for use as bottom and top layers as labelled on the pictures. Scale bar is 25 μm.

In general, oxidized silicon substrate ($SiO_2$/Si) is used for exfoliating 2D materials. The $SiO_2$ layer on Si substrate is known to enhance the optical contrast of graphene to enable the identification using optical microscope [91]. Optical contrast is developed as a fast and reliable way to estimate the layer thickness of 2D materials. In addition, other methods such as atomic force microscopy (AFM) and Raman spectroscopy can provide an independent verification of the flake thickness [91]. The size of graphene flakes on the $SiO_2$/Si substrate can be as large as few millimeters depending on cleavage plane, size and flatness of the bulk 2D crystal. Monolayers or thinner flakes of hexagonal boron nitride (hBN) can also be produced by using this exfoliation method and can be identified on 70 nm $SiO_2$/Si wafer with better contrast for monolayer of hBN [92]. Similarly, $MoS_2$ and mica thin flakes can be produced on the $SiO_2$/Si substrate as well. Mostly we use graphene, hBN or $MoS_2$ flakes as top, bottom and spacers for fabricating Å-channel devices. The optical images of 2D materials such as graphene, $MoS_2$ and hBN, which are used to prepare top, bottom and spacers are shown in Fig. 6.

*Spacer design and fabrication (Step 31-49)*

Typically, graphene layers of different thicknesses are mechanically exfoliated on to a 290 nm $SiO_2$/Si substrate. On the flake of desired thickness, EBL is done to pattern an array of parallel stripes which are subsequently etched using oxygen-argon plasma to create graphene spacers. Width of the channel is designed to be between 50 nm to 130 nm, separated by a spacer of ~170 nm to ~1 micron in width. The substrate carrying the desired graphene flake to be used as a spacer is coated with an e-beam resist. In our case, PMMA (polymethyl methacrylate - 950K molecular weight, 3% wt/vol in anisole solvent) is used as an e-beam resist with a thickness of 140 ± 10 nm. Then patterning is carried out on EBL to write the spacer's design with an area dose of 210 μC/cm$^2$.

*Taking the stage coordinates (Steps 34-35).* Alignment marks are necessary for locating the desired flake on the substrate to write with EBL. We use a ProScan™III motorised stage system, in which coordinates are taken by assigning two marks on the horizontal corners of substrate (Fig. 7a). The stage coordinate system comes with a software and coupled to microscope used for recording the flake position on the substrate. Flake coordinates are recorded when the stage is at the flakes' exact location. It is possible to record multiple flake positions on the same substrate for writing alignment marks using EBL.

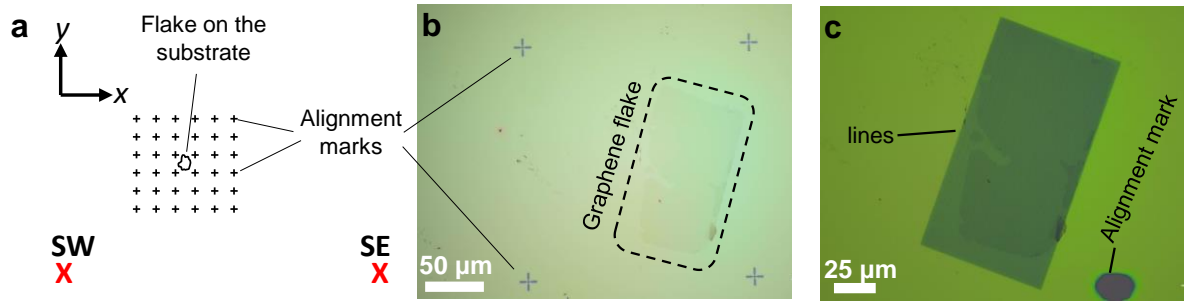

**Fig. 7| Fabrication of spacers by using electron beam lithography (EBL).** Using PMMA as a resist, two rounds of EBL are performed: the first to pattern alignments marks near each flake and the second to write lines design on the spacer flake. **a,** Example of coordinate system used for EBL. Two red crosses at the bottom corners of the substrate are used to set up global coordinates to locate flakes, and alignment marks are used for accurate local alignment relative to flakes for second EBL. **b,** Optical microscope image of a bilayer graphene (in dashed outline) next to alignment marks after first EBL. **c,** Optical microscope image of the same bilayer graphene with lines written in the second round of the EBL. These lines on PMMA can be used as a mask for dry etching through the graphene to produce the desired spacers.

*Nanochannels creation by EBL (Steps 36-43).* In EBL, a focused electron beam is scanned over a substrate covered with an electron-sensitive polymer layer. As PMMA is a positive resist, the irradiated areas can then be removed using a developer. Commonly used developers are mixture of methyl isobutyl ketone (MIBK) and IPA (1:3 vol/vol) or water and IPA (1:3 vol/vol, cooled down to 5 °C).

The EBL setup includes a scanning electron microscope and a pattern generator. The coordinates recorded using optical microscope attached ProScan™III are entered in to patterning software, Elphy quantum system to perform translational and angle/tilt correction in global coordinates. Once the beam current is stable and measured, it is important that the focus and stigmation are corrected and maintained. Alignment marks (e.g., crosses) are first designed in the Graphic Design System (GDSII) format, and then drawn in a pattern extending to 1000 µm in ±X and ±Y direction with a pitch distance of 200 µm between them.

Post exposure, crosses were developed in MIBK:IPA (room temperature) or water:IPA (cooled to 5 °C) mixture for 30 s and immersed in IPA for 30 s to stop further development and dried with nitrogen gas. After development, images of the flakes are taken at 5×, 10×, 20× and 50× magnification and embedded in software like Inkscape or Layout editor and are precisely aligned. The desired pattern to create channels is drawn on the flake, e.g., 100 nm wide stripes spaced by 170 nm are drawn onto the flake and the file is saved to be read out by patterning tool (Elphy Quantum on Raith) for exposure.

We use a write-field of approximately 250 µm × 250 µm to pattern the nanochannels. Once global coordinates are set, the stage is driven to the flake position. Upon switching to local coordinates, set the flake coordinates as origin. From the flake position, we search within a square of ± 500 µm above and below to locate the position of alignment marks to correct and align the angle (Fig. 7). The line pattern is then exposed by e-beam, and developed. It is important to use the optimized development conditions, as longer development will result in merging of lines leaving a rectangular box or overdeveloped channels after etching. After the development step, the substrates are blow-dried with a nitrogen gun.

*AFM confirmation of EBL pattern (Step 44):* AFM analysis is carried out to find the width and height of the lines developed on PMMA and to ensure that lines are exposed properly and are thick enough to sustain the etching step**.** After the development step, the PMMA layer swells, so we wait for 5 to 6 hours at room temperature to do the AFM, or one can also bake it at 80 °C for 5-10 min to evaporate the solvent from PMMA. With our spinning recipe (Table 1), PMMA layer thickness is about 140 ± 10 nm. In an ideal condition, the channel and spacer line width will be commensurate with the design used.

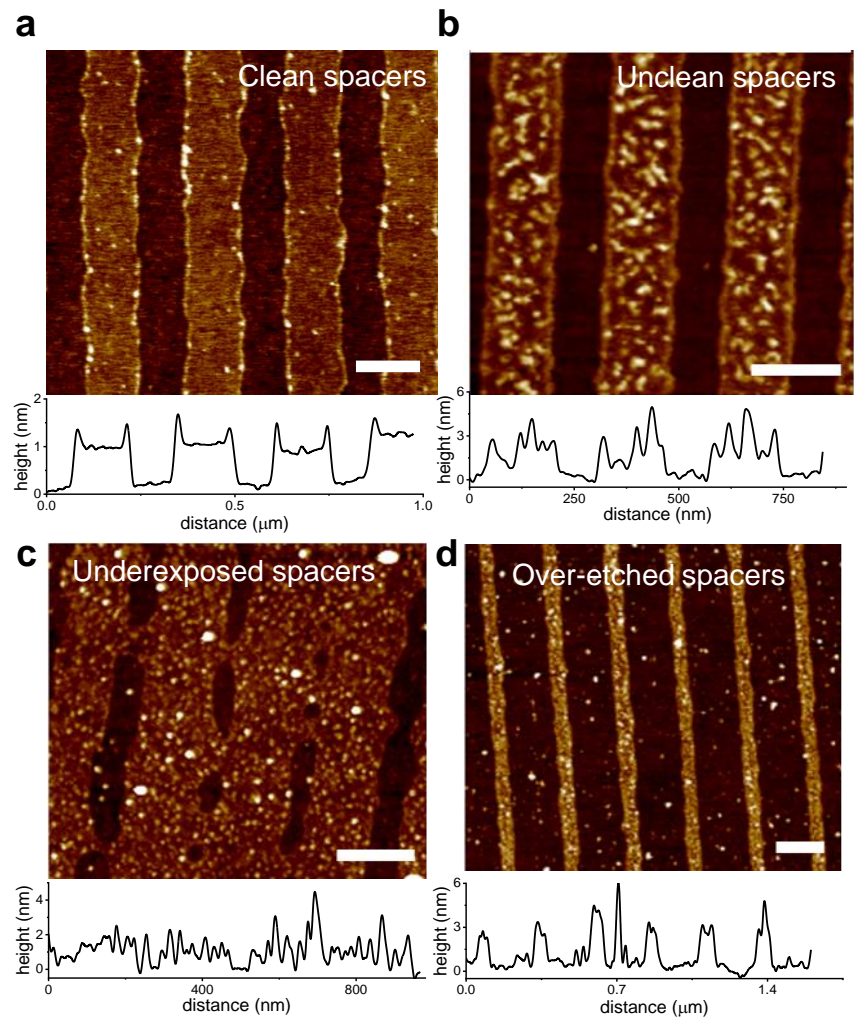

**Fig. 8| Atomic force microscopy (AFM) images of spacers.** The AFM images and height profiles of **a,** cleaned graphene spacers of height ≈ 1 nm; **b,** uncleaned graphene spacers with PMMA residue; **c,** graphene spacers showing cross stitching that could be due to under exposure during EBL and/or insufficient etching time; **d,** graphene spacers which are over-etched resulting wider channels and narrower spacers. Scale bars, 200 nm.

**Influence of focus and dose.** EBL can result in slightly different dimensions of the channel than the above-specified range due to improper focus leading to inadequate exposure, which can either result in underexposed or overexposed patterns. AFM image of correctly exposed and underexposed lines are shown in Fig. 8. With the design being 100:170 nm channel to spacer ratio, an optimized e-beam exposure will result in PMMA line depth of ~140 nm, channel to spacer width of 100:170 (nm) and in comparison, underexposed PMMA line depth is found to be < 70 nm channel to spacer width 70:200 (nm). Although PMMA acts as a mask, the plasma etches both transversely and laterally causing the channel to widen.

*Etching of graphene channels (Step 45).* Spacer flakes are dry etched, with PMMA serving as a mask in the regions where it is not exposed to the e-beam. For instance, soft etching recipe of graphene is done with recipe R5 (Table 1), where the etch rate is approximately 20 s per 1 layer of graphene. Plasma exposure is carried out step-wise (steps of 20 s) for more than one layer of graphene to avoid hardening of the PMMA, as hardened PMMA is difficult to clean in later steps. It is to be noted that in the etching process timing is crucial as an over-etch or under-etch can result in wider and narrower channel widths causing the top and bottom layers to either sag or result in closed channels.

*Cleaning (Steps 46-49).* After the graphene etching, we remove the PMMA mask by cleaning with acetone and IPA. Nanochannel devices require clean spacers for many reasons : (i) Localized polymer residues increase the channel height and cause non-uniform channels; (ii) Molecular and ion transport can be affected by residue accumulation inside channels, resulting in a decrease in conductance; iii) As a consequence of uneven thickness due to the presence of PMMA residues, the top crystal can delaminate when even small voltages are applied during ion flow measurements. The cleaning protocol is divided into two parts. The first step involves washing the PMMA polymer residues with mild sonication using recipe R7 (Table 1) by moving the substrate in and out of acetone every 10 seconds. In the second step, remaining PMMA is removed several times with hot acetone at 50°C for a few hours, followed by IPA rinse and nitrogen blow drying. This two-step cleaning protocol is repeated if the PMMA is not completely removed. All the cleaning solvents were used directly from the bottle (purity, 99.8%) without any further purification. In order to check if the spacers are clean, they are scanned using an AFM after cleaning procedure (Fig 8). If the spacers are still not clean, an annealing of spacers to 400 °C for 4 hours can also be carried out to remove further PMMA by thermal degradation.

A comparative AFM images of cleaned, uncleaned, under-exposed and over etched lines with their respective height profile are presented in Fig. 8. All these lines are made on a tri-layer graphene flake. Clean lines in Fig. 8a gives around ~1 nm thickness of spacer whereas uncleaned lines (Fig.8b) are rough with more than double the thickness. The AFM image in Fig.8c shows the lines which are under exposed and cross-stitched while Fig. 8d shows the over-exposed lines with channel and spacer widths 200 nm and 70 nm respectively. These lines (showed in Fig. 8b, c and d) are not good for use as spacers as the resulting nanochannels after the top flake transfer could merge, collapse and sag during the device fabrication or result in clogged devices. Over-etching and cross-stitching can occur for many reasons, including unstable beam currents, poor focusing, stigmatism correction, over/under dose, and etching time. These parameters must be optimized for making clean spacers.

*Transfer of flakes (Steps 50-80)*

Custom-made micromanipulators underneath an optical microscope are used in the transfer setup which consists of five main parts: temperature control panel, optical microscope, transfer stage, mechanical transfer arms and vacuum switches as shown in Fig. 9. The transfer stage temperature is typically between 50 °C to 90 °C for transferring the flakes. The substrate and arm are held by the vacuum.

There are three well known transfer methods which can be used to produce van der Waals heterostructures or stacks of 2D materials, (1) chemical etching based wet transfer, (2) PMMA/PMGI dry transfer method and (3) PDMS/PPC dry transfer. For fabricating our Å-channel device, usually wet transfer technique is used to transfer flakes because of reliability.

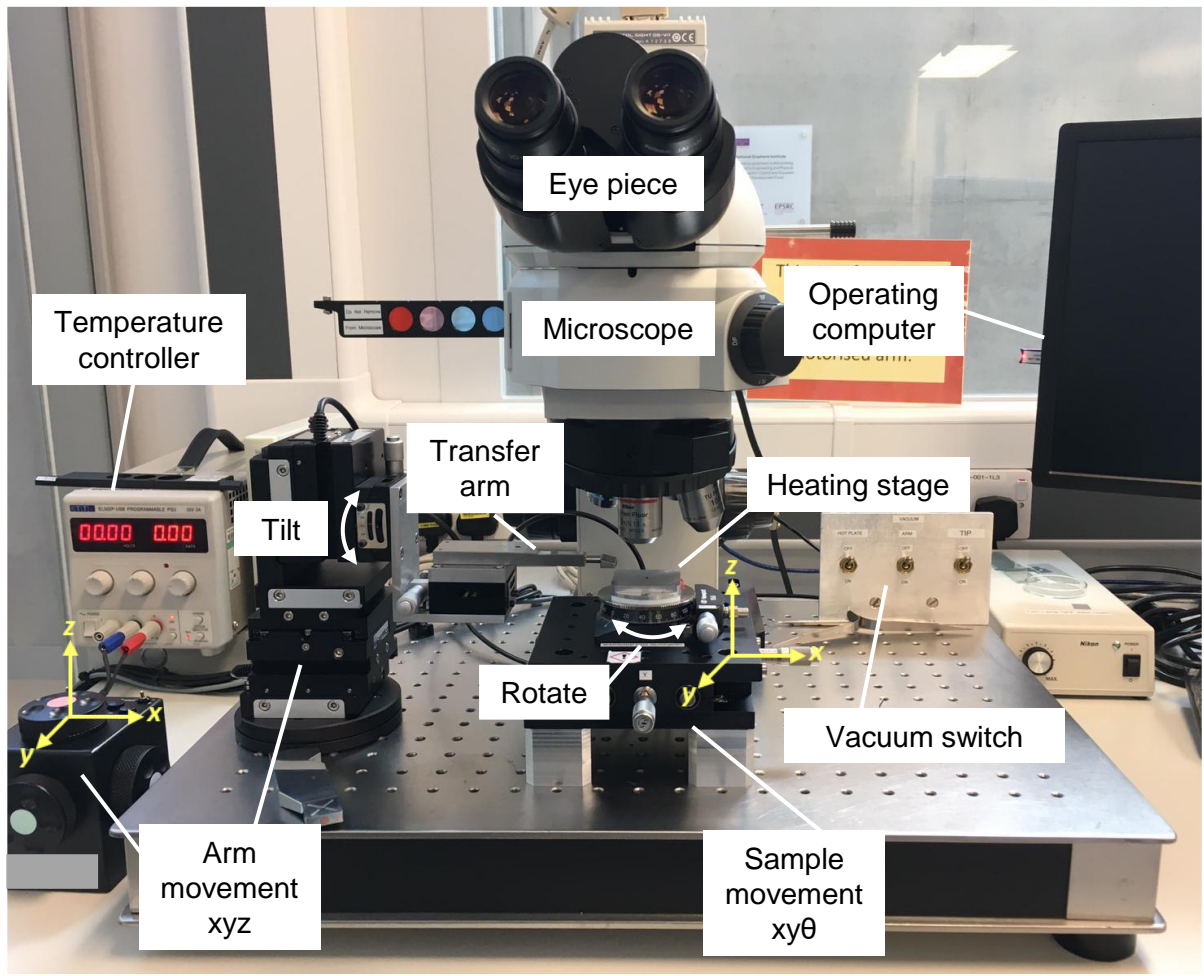

**Fig. 9| Photograph of in-house flake transfer setup.** The transfer arm has three translational degrees of freedom and the substrate stage has two translational and one rotational degrees of freedom. The transfer arm is motorised and controlled by the arm movement controller on the left. An optical microscope allows for accurate alignment, and a heating stage is used to heat the sample/substrate, both controlled by the operating computer on the right. The vacuum line keeps the substrates secured to the sample stage and transfer arm.

**Wet transfer.** The wet transfer method uses a direct contact with dilute KOH solution and deionized (DI) water (Fig. 10) to lift the flake off the substrate, which can then be transferred onto another desired substrate. It is one of the standard methods to transfer flakes from one substrate to another or build van der Waals heterostructures [93]. The substrate with flakes to be transferred is first spin coated with PMMA with the recipe R6 (Table 1). Subsequently the wafer is immersed in dilute KOH solution (~10% wt/vol) to etch away silicon oxide slowly until the PMMA membrane carrying the flakes floats in the solution. Typical etching time for this process is ~6 to ~10 hours. The floating PMMA membrane is carefully washed by dipping in DI water (not sprayed) and then the flake is ready for transferring to a target substrate. It is worth noting that one can accelerate KOH etching process by using higher concentrated KOH solution at higher temperature. However, we use dilute KOH solutions and room temperature (20 °C to 30 °C) to avoid contamination risk from high concentration KOH and tape melting. Bottom and top flakes can also be transferred using dry transfer technique [52].

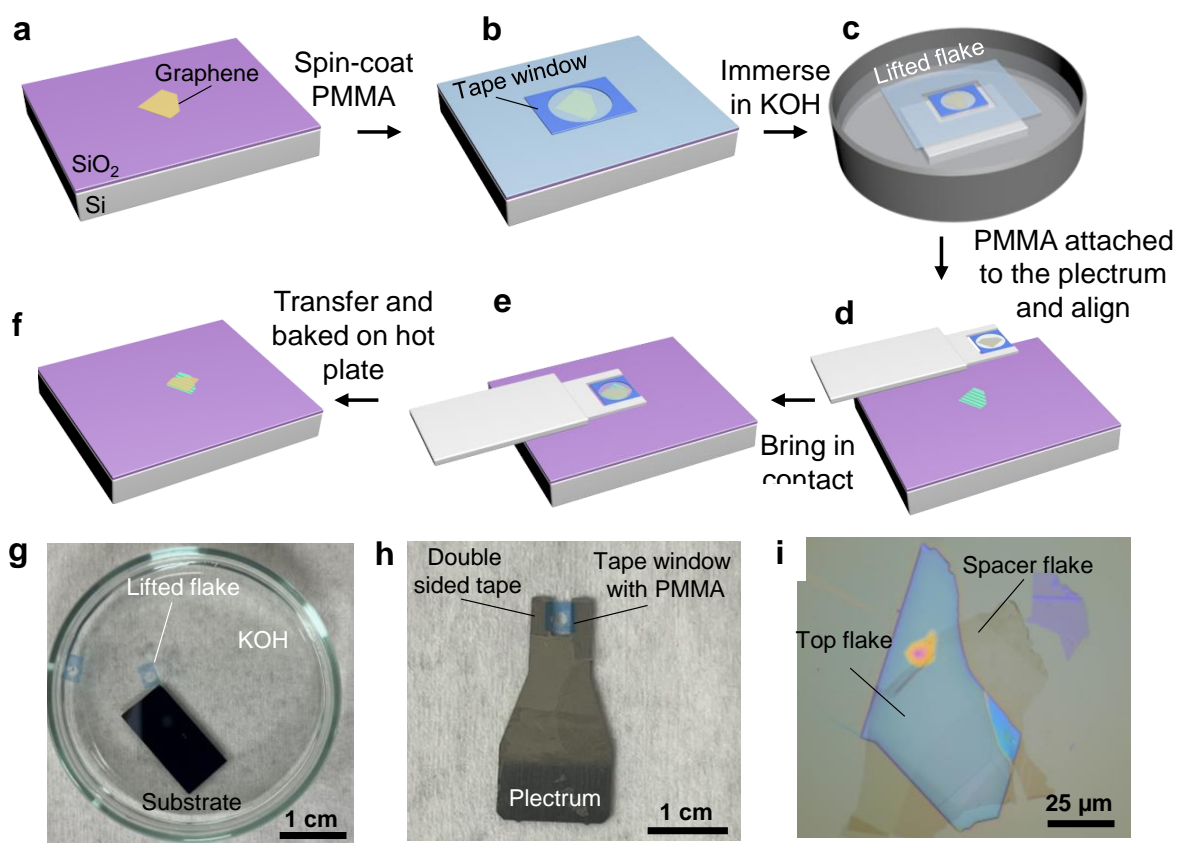

**Fig. 10| Schematics of wet transfer process. a,** Exfoliation of graphene on SiO$_2$/Si substrate. **b,** PMMA is spin-coated on the substrate, baked, and then a tape window is put on PMMA around the desired flake. Once the tape is firmly adhered to the PMMA film, the edges of the tape are scratched to remove the PMMA and provide access to KOH in the next step. **c,** The substrate is then immersed in aqueous KOH for selective etching of the SiO$_2$ and detachment of the PMMA/flake supported by the tape **d,** The tape window is scooped out from the DI water and placed on plectrum. Under a microscope, the plectrum is mounted on the transfer arm, and the flake on PMMA is aligned with the target substrate. **e,** The membrane is then brought in contact with the target substrate. **f,** The PMMA is removed after the post bake and by gently washing with acetone. **g,** Photograph of the substrate in dilute KOH solution in a petri dish. The tape window carrying PMMA/flake floating in the KOH solution. **h,** Photograph of plectrum carrying PMMA membrane with flake. **i,** Optical image of the transferred graphene spacers on the MoS$_2$ flake.

**Optical alignment and flake transfer.** The flake with a PMMA membrane is attached to double-sided sticky tape on the plectrum which is held by vacuum on to the transfer arm allowing independent motion in x, y, and z directions. The target substrate is held by vacuum suction on a temperature-controlled stage. After the flake and targeted flake/substrate are aligned, the transfer arm with the plectrum moves down until the target substrate is in proper contact with the PMMA membrane. Increasing the temperature of heating stage will influence the adhesion between the flake and the substrate for a successful flake transfer. Our homemade micromanipulator comprises an optical microscope which is connected to a computer to visualize the transfer process from the microscope camera. We draw contours on the flakes to help in the alignment as the flakes move out of focus while lowering the membrane to bring in contact with the substrate. By using an optical microscope with filters and different objective lenses, flake visibility can be improved for better flake alignment while performing transfers [94].

*Device post-processing (Steps 81-107)*

Prior to the use of the device, few post-processing steps should be carried out to ensure good functionality and reliability of the final device. During Annealing (Step 81) **it** is crucial to remove contamination after each transfer process, such as adsorbates from the atmosphere or polymer residues from fabrication. After fabrication, a device needs to be annealed in 10% $H_2$ in Ar atmosphere for a few hours, and the annealing temperature is determined by the thermal stability of the selected 2D materials. For instance, the devices made of graphite or hBN crystals can be annealed at 400 °C while the annealing temperature should be decreased to 350 °C in the case of $MoS_2$ channels. Besides removal of fabrication residues after each flake transfer step, annealing should routinely be applied before every use of the final devices. Graphite and hBN devices are annealed at 300 °C for 3 hours and 400 °C for 4 hours, whereas $MoS_2$ channel devices are annealed at 300 °C for 1 hour, followed by 3 hours at 350 °C under the flow rate of 10% $H_2$ in Ar gas mixture.

In the fabrication, the top crystal is carefully chosen with sharp edges for obtaining open and stable channels. However, a clogged channel is still a possibility during the device fabrication as it is common to have a step-like edge on the thick top crystal flakes. In our experience, a top crystal thinner than 70 nm can sag into the spacer and if the crystal is much thinner (< 20 nm) it can completely block the channel [57]. To solve this problem with thin edges, we use lithography to pattern and deposit gold film above the top crystal (the detailed process is shown in Fig. 11). Typically, the gold patch is deposited in a strip-like design, which is aligned along the hole of $SiN_x$ membrane.

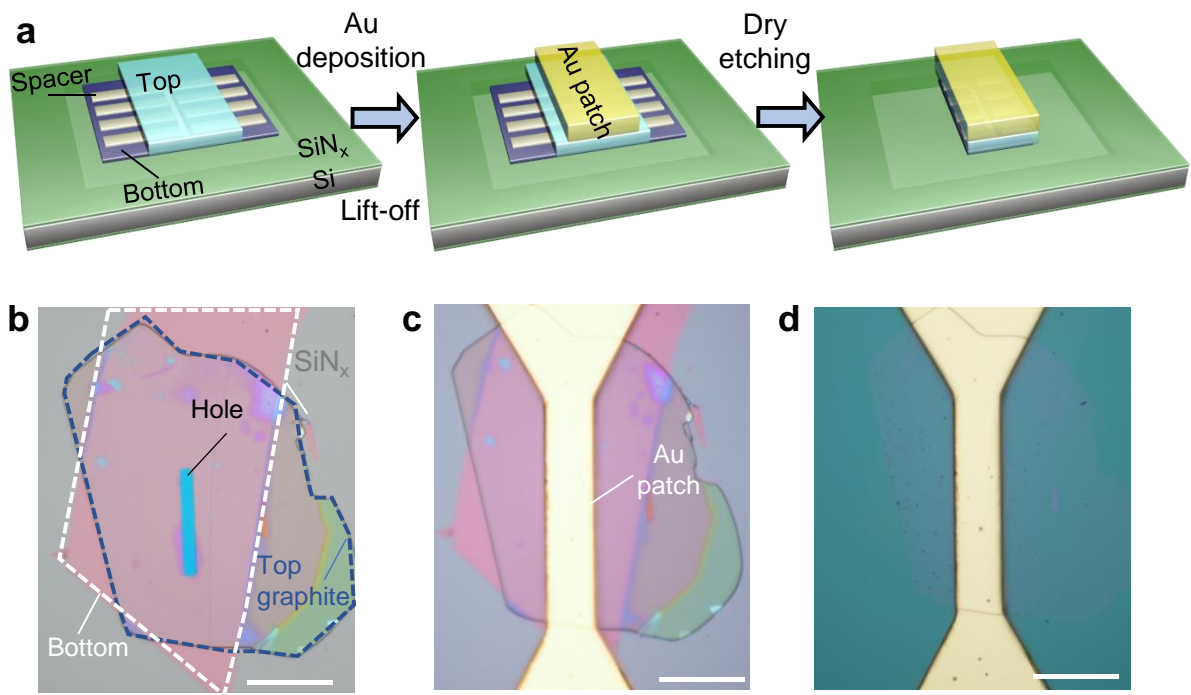

**Fig 11| Device post-processing to open channels *via* gold patch deposition. a,** Schematic process flow for opening channels and defining channel length. Starting with a typical channels stack on a $SiN_x$ membrane, photolithography and metal deposition are used to create Au patch on the device. Dry etching opens the channels. Optical image of a typical capillary device in reflection mode **b**, before and **c**, after the gold patch deposition and lifting process, and **d**, after etching. The gold patch is visible in yellow colour in b, and c. In a, the bottom flake is highlighted with white-coloured dashed contour and top flake is highlighted as blue dashed lines. Scale bar on all optical images, 20 μm.

One can use either photolithography or electron beam lithography to write the gold patch design. The deposition of gold patch (5 nm Cr layer followed by 50 nm Au layer) is done via physical vapor deposition (e.g., e-beam evaporator, thermal evaporator or sputtering). A lift-off technique is used to remove the unwanted gold layer on the device by soaking in hot acetone (~ 50 °C). This gold strip can act as a mask in the next step of RIE etching to open the channel entries and control the channel length. Optical images of the capillary device before and after the gold patch deposition are shown in Figure 11b, c and d.

Advantages of patterning gold patch on 2D channel devices are: the prevention of sagging of the top layer crystal edges into the channel, definition of channel length, control of the number of open channels on the device, mechanical support to the device by protecting it from structural damage, during the flow measurements in liquid environment.

**Characterization of the device**

All the fabricated channel devices can be evaluated by helium leak test (discussed in "Experimental Setup") or by performing ion- flow measurements. Generally, if a device shows higher ion conductance (>2 times) or higher helium flux (>three orders of magnitude) than estimated values based on the channel dimensions, it will be considered as 'delaminated device'. If a device shows lower flux or conductance (by more than an order of magnitude), annealing step will be repeated to regenerate the device.

**Storage of the device**

After performing measurements, the device can be stored in water or > 90 % relative humidity (RH) environments, usually for a month. We store our devices in activated charcoal to increase the lifetime (few months to several years). Such stored devices are annealed before reuse in measurements to restore the previously recorded conductance or flux [95].

**Experimental Setup**

**Gas flow setup (Steps 108-118)**

A mass spectrometer (Leybold phoenix quadro dry, Helium gas leak detector) with mass range 2 to 4 is used to study the gas transport behavior through these Å-channels. A custom-made sample holder was designed to hold our Å-channels devices between two conical brass compartments with O-rings (leak-tight) on both sides (Fig. 12). These two compartments are clamped together with help of screws and bolts and separating entry and exit of Å-channels as schematically illustrated in Fig. 12b. The Å-channels device, which is usually on a SiN$_x$/Si substrate with ~17 mm × 17 mm or 12 mm × 12 mm in dimensions, is loaded on the holder as shown in Fig. 12 with appropriate diameter of O-rings.

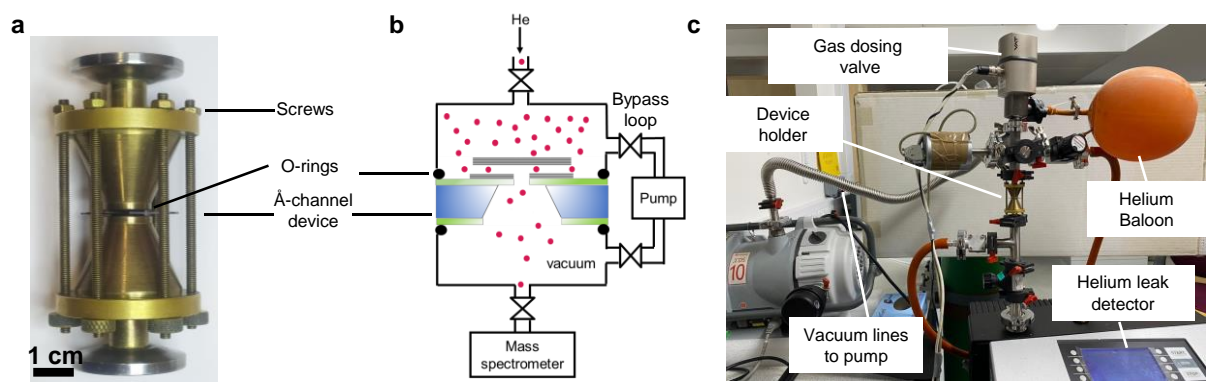

**Fig. 12| Gas flow measurement setup. a,** Photograph of a sample holder. A pair of two conical shaped compartments made up of brass are clamping the wafer chip/Å-channels device with two chemical resistant O-rings and adjustable screw fittings. **b,** A schematic representation of the gas measurement

setup. The Å-channels device (sandwiched in the sample holder) is mounted between a feed compartment and a mass spectrometer. The feed and permeate compartments are evacuated using a vacuum pump with both connected using a bypass vacuum loop. **c,** Photograph of gas flow measurement setup.

The two compartments of the sample holder are connected to the top and bottom chambers on the measurement setup, one above and another below the sample holder. The top chamber, i.e., feed chamber is equipped with a gas dosing valve (which connects a He gas balloon to feed chamber), and a pressure gauge, whereas the bottom chamber is connected to a mass spectrometer (He leak detector). These both chambers are connected to a vacuum pump *via* valves, and a bypass loop. The Å-channel device is loaded on the custom-made brass holder and mounted between both the compartments. The screws on the brass holder are gently fastened to keep the device in the position and well aligned between O-rings. Both the chambers are evacuated to <1 mbar using bypass vacuum loop (Fig. 12b and c). Helium gas is introduced through the feed (top) chamber to flow through Å-channels on the $SiN_x$/Si chip, into the mass spectrometer through the bottom chamber. The gas dosing valve (operated by voltage) is used here to release the helium in a controlled fashion. The leak detector gives the quantitative leak rate of the He flow (Fig.16). The brass sample holder is first tested to check if it is leak-tight by loading a blank $SiN_x$/Si chip without any channels or hole.

**Water flow measurement (Steps 119-125)**

A custom-designed gravimetric sample chamber is used for the water flow experiment. It composes of a container and lid made from aluminium where the lid has an opening above the capillary device that is the same diameter as the container's opening (10 mm) as shown in Fig. 13a. There are grooves on both container and lid surfaces which are designed to fit O-rings. Our device is placed in between the container filled with water and the lid which are fitted with the O-rings. These two counterparts can be attached together through the aligned screw positions. The well-fitted O-rings in grooves can exert a tight clamping force to hold the device in a stable position, and the water flows only through the channels in the device without any leaks. When mounting the device on the container, the silicon wafer edges should not cross any of the screw holes at the container/lid edges. One can easily create a 100% relative humidity (RH) environment by filling the DI water into the container, or the container can be flipped to make the water (liquid) in contact with the channel entries. Importantly, this gravimetric sample holder chamber can be made from a variety of materials not restricting to aluminium, e.g., brass, polyether ether ketone (PEEK), corrosion-resistant alloys etc., as long as the total weight of the container is not above the maximum weight limit of the balance.

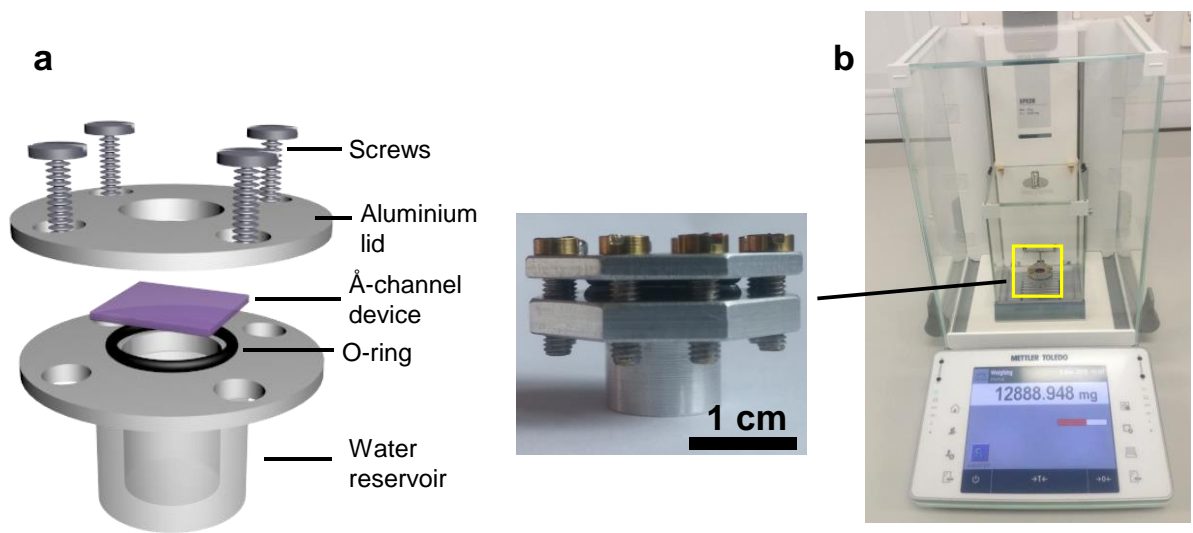

**Fig. 13| Gravimetric sample holder and measurement setup. a,** Components of the gravimetric sample holder made out of aluminium. A small reservoir is filled with water (0.5 mL) and covered by a SiN$_x$ substrate with a capillary device. With the well-sealed O-rings, this gravimetric cell can be used to measure flow rates through Å-channels, as the water can only evaporate from the sample holder *via* the channels. On the right is the photograph of a closed cell. **b,** Photograph of our gravimetric set-up which shows microbalance with the gravimetric cell (yellow square).

The gravimetric measurement can be done in two ways using the sample holder (Fig.13a), either vapor-in-contact or water-in-contact with the channels. For vapor-in-contact, the Å-channel device faces the chamber, only interacting with the water vapor, whereas for water-in-contact, the device is directly contacting the water liquid. In either way of measurement, the amount of water which has flown through the channels is the gravimetric loss i.e., weight of the sample holder including water which is monitored by a microbalance with a high sensitivity and accuracy of 1 µg (Fig. 13b). The set-up with the device in between is simply placed in the microbalance. As the microbalance is sensitive to temperature changes, we use an environmental chamber to maintain the temperature and humidity. We used a constant temperature and a set relative humidity (e.g., 20 °C, 30% RH) for our measurements. For mounting the samples, either sides of the device i.e., the channels side or the micro-hole side of the Si/SiN$_x$ chip facing the chamber were tested by flipping the device before mounting, and we found that both ways give similar flows.

**Ionic current measurements (Steps 126-144)**

The Ag/AgCl electrodes were used in ion transport measurements. Such electrodes are non-polarizible, simple to prepare, and promote a stable electrode potential. Our Ag/AgCl electrodes are home-made and prepared by immersing the Ag rods (2 mm in diameter and 3 cm long) into sodium hypochlorite (NaClO) overnight. Each set of electrodes typically lasts for several device measurements up to a month. The electrodes are then thoroughly rinsed in DI water and dried. Before the measurement, a fine sandpaper is used to expose the pure silver on the extreme end of the electrode. For those ion transport measurements that could promote redox reactions on the electrodes generating undesired offset voltages, we use commercial saturated salt bridge electrodes (CH Instruments, USA).

Aqueous salt solutions for the ionic measurements were prepared from a 1 M stock solution of the salt of interest in deionized (DI) water. Once the desired concentrations are established, dilutions were made accordingly, down to $10^{-6}$ M.

A custom-made electrochemical cell machined from chemically resistant PEEK is used to measure ion current through the fabricated devices (Fig. 14). Our customised electrochemical cell has the role of securing the device while completely sealing the chip to avoid leakage currents. Overall, it consists of two reservoirs with equal volumetric capacity. The fabricated devices are placed between the reservoirs and sealed using acid resistant O-rings (6 mm or 11 mm in diameter) and a set of screw and nuts. The electrolyte solution is added into the cell through the holes on the top. The volume of reservoirs is relatively large and the cell design is open to air, which prevents the formation of air bubbles during the filling of solution. The cleaning the cell after every measurement is important to completely removal of residues from previous electrolytes. In our case, the cell is immersed in a mixture of IPA and DI water (1:1 ratio) and sonicated for 1 hour, followed by repeated washes of water, IPA rinsing and finally blow dry using nitrogen air gun.

A Keithley 2636B source meter is used for nanofluidic measurements due to its versatility, good stability and noise performance. The source meter has a measurement resolution of 0.1fA/100nV. Our measurements are usually done at low frequency and have low current output (in the nanoampere range). We use a *LabVIEW* 2018 version program to conduct ionic measurements and collect data for the characteristic *I-V* curves, drift-diffusion measurements to track the time-resolved diffusion current across the electrodes with zero applied voltage. Once the current is saturated, *I-V* curves (drift) are measured with voltage range of ±0.2 V to ±0.5 V with step size of 2 mV.

Our voltage-pressure cell is custom-made and machined from PEEK material. This design is similar to the electrochemical cell described above, except that the two reservoirs with equal volumetric capacity bear inlet and outlet on their outer sides (Fig. 14c) to be connected to the pressure pump.

Owing to its fast (50 ms) and accurate (± 1 mbar) pressure control, a microfluidic pump (Dual AF1, Elveflow) is used for streaming current/potential measurements. The system is operated by Elveflow Smart Interface (ESI) software with accessible pressure varying between -700 mbar and +1 bar with step size 1 mbar.

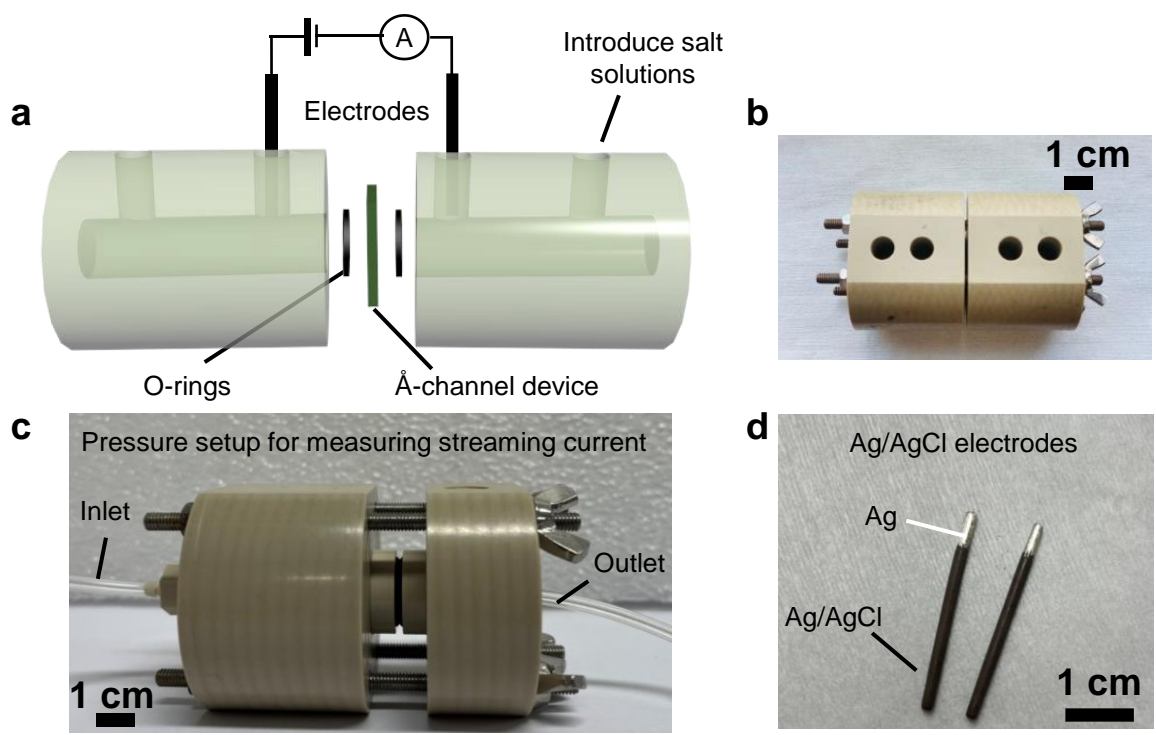

**Fig. 14| An electrochemical measurement setup. a,** A schematic of the electrochemical cell made from PEEK used to investigate ion transport using a two-electrode system. Salt solutions can be added *via* the holes on the top of a cell, whilst the device can be mounted in the centre of the cell between the O-rings on the two central holes. Electrodes can be connected *via* any one of the top holes of the reservoirs. **b**, Photograph of the electrochemical cell. The screws that hold the two cells compartments are visible. **c**, Photograph of streaming current/voltage measurement cell which shows feed (inlet) and permeate (outlet) valve for pressure. Both the compartments hold volume of 2 ml each. **d**, Photograph of Ag/AgCl electrodes.

**Data acquisition**

To validate the efficacy of the measurement setup, we first perform the measurement on several $SiN_x$ membranes each containing a circular aperture with various diameters. For example, in the case of gravimetric measurements using the same set-up and measurement conditions, the obtained water loss as a function of the aperture area is linear [57]. This slope represents the water evaporation rate, and it is consistent with the theoretically estimated flux values from the $SiN_x$ micro hole [57]. Thus, gravimetric measurement is a reliable technique for investigating the water transport behavior.

LabVIEW (National Instruments) offers an easy-to-use environment, allowing the acquisition program to be quickly developed for a wide range of applications, including electrochemical and ion-transport measurements [58,60,61]. Nevertheless, other programming platforms such as Python, are good alternatives with sufficient flexibility.

A home-built setup is used to measure the gas permeation through the devices. A leak detector (Leybold phoenix quadro dry) is used to measure the permeation rate of the gas passing through the device with a controlled pressure. The input pressure is controlled by a gas dosing valve connected to the gases and input of the sample holder. Gases are fed from the balloon into the input chamber, which has its pressure monitored using an attached pressure gauge. Gas rate is set by a controller to a gas dosing valve. The rate of gas flow is then measured through the leak detector attached at the output end, so that any pressure changes and leak rate values are continuously transmitted to a computer. The values are recorded using a LabVIEW-based program.

By keeping the set-up on the high accuracy weighing scale connected to a computer through an interface program of choice, the gravimetric loss against over time up to several days can be recorded for all devices with different channel dimensions. For our gravimetric measurements, we use a LabVIEW program for data acquisition and recording. The measurement of each device is repeated at least three times to ensure the reproducibility of the data.

Software improves the workflow during the experiment, making the measurements efficient and reproducible. A wide range of software tools or graphical programming environments with high flexibility are available online. As such, the software is recommended to be user-friendly so that all measurement parameters can be selected and/or tuned during the measurement. In our experiments, the source-meter is controlled by a LabVIEW-based software which allows us to input and control the following parameters:

- Voltage range: source voltages at which the *I-V* curves will be acquired, typical voltage lies between -200 mV to +200 mV.
- Voltage sweep comprises the linear variation of the voltage, this can be adjusted in the LabVIEW / interface program, and our typical sweeping voltage is 2 mV/s.
- For each measurement condition, we can specify the number of cycles to repeat *I-V*. Using a separate program, time-resolved current measurements (diffusion, at zero voltage) are performed to monitor the saturation.

The software automatically generates the files from the acquired *I-V* or *I-t* curves in a general format (e.g., .lvm, .txt, or .dat), so that it can be read by common data-analysis software, such as Origin, Microsoft Excel, MATLAB etc.

**Data analysis**

The measured gas flow output data files contain pressure change and leak rate along with the time sweep. Pressure change is plotted against leak rate using data plotting software (e.g.; Origin 2021b, Microsoft excel, data plot etc.). Leak rate varies linearly with the pressure for open channels. The slope of this curve is obtained by a linear fit. This slope is the gas permeance value and it varies corresponding to the channel height, number of channels and channel dimensions. This permeance value can then be used to estimate the channel size if we know all the channel dimensions.

From the gravimetric measurements, we acquire accurate time-resolved weight changes $|U-U_0|$ of the cell (mounted with Å-channel device), where $U$ and $U_0$ are the final and initial weight, respectively. Thus, the time-resolved weight loss is assessed by plotting the weight loss over the respective time interval. Such plot is linear, and its slope gives the water evaporation rate.

The recorded data file from the measurements contains current and voltage data along with time stamp. The slope of this curve corresponds to the experimental conductance ($G_{exp} = 1/R$). In this way, $G_{exp}$ of a desired salt is measured for different concentrations of electrolyte solutions. The expected conductance ($G_{bulk}$) is estimated using equation 1.

$$G_{bulk} = \sigma_{bulk} \left(\frac{A}{L}\right) \qquad (1)$$

where, $\sigma_{bulk}$, A and L are the bulk conductivity of electrolytes in aqueous solution, cross-sectional area of the channels and channel length respectively. Area of the respective device is determined by the channel dimensions such as height, width and number of the channels. For very dilute solutions, to calculate an equivalent conductivity ($\Lambda$; units $10^{-4}$ m$^2$ S/mol) with respect to the concentration C of the solution, Debye-Hückel-Onsager eq. 2 [96] was used: The equation is reliable for $C < 0.001$ mol/L; with increasing concentration the error will be higher.

$$\Lambda = \Lambda° - (A + B\Lambda°) \sqrt{C} \qquad (2)$$

where $\Lambda°$ is the equivalent (molar) conductivity and diffusion at infinite dilution can be found in CRC hand book [97]; A and B are the 'constants', e.g., for monovalent ions the values are $A = 60.20$ and $B = 0.229$ for $T = 298$K [97]. The conductivity of the solution σ in S/m is calculated by multiplying $\Lambda$ with C ($\sigma = \Lambda \times C$, for monovalent ions). There are more complex treatments available to calculate the molar conductance for higher concentrations [96] as well as tabulated values for certain solutions [97].

To find the mobility μ (noted as $\mu^+$ for cations and $\mu^-$ for anions), the recorded file of diffusion experiment (discussed in experimental setup section) are also plotted in the similar way as in conductance measurement. The I-V curve will give the value of liquid-junction potential, $E_m$ (or zero-current potential, value of V for I=0) arising from different mobilities of cations and anions in bulk solutions. The mobility ratio $\mu^+/\mu^-$ were determined by Henderson formula [98] eq. 3:

$$\frac{\mu^+}{\mu^-} = -\left(\frac{z_+}{z_-}\right)\left(\frac{\ln(\nabla)-z_-FE_m/RT}{\ln(\nabla)-z_+FE_m/RT}\right) \qquad (3)$$

where, $z_+$ and $z_-$ are the valence of cation and anion respectively, F is the Faraday constant, R is the universal gas constant, $\nabla$ is the ratio of feed and permeate concentrations in the containers and T is the temperature in Kelvin. Individual mobility can be calculated by the conductivity which can be described in terms of ion mobilities as $\sigma \approx F(z_+C_+\mu^+ + |z_-|C_-\mu^-)$ where $C_+$ and $C_-$ are the concentrations of anions and cations, respectively.

To obtain the electro kinetic mobility ($\mu_{stream}$) values from the pressure-driven currents are calculated the slope from streaming current per channel versus applied pressure per length. The slope of $I - P$ curve can be described as $\mu_{stream} = I_{stream} / (NA\Delta P/L)$, where $I_{stream}$ is streaming current, N is number of channels, A is area of the channel ($wh$), $\Delta P$ pressure gradient, L is length of the channel. From this equation, $\mu_{stream}$ can be obtained.

A variety of software tools are available for the data treatment discussed, which is mostly comprised by curve fitting or regression of the experimental data. Therefore, the general requirements is first to install an efficient fitting package, such as Origin Pro, Igor Pro, Mathematica, Maple, or MATLAB.

**Materials and reagents**

- 4-inch silicon substrates (orientation: <100>, 1-side polished, doping: p-type (Boron), resistivity: 0.001-0.01 Ω cm, thickness: 525 ± 25 μm, MicroChemicals GmbH, Cat. No. WSD40525250B1050SNN1)

- 4-inch silicon substrates with 90 nm silicon oxide layer (orientation: <100>, 1-side polished, doping: p-type (Boron), resistivity: 0.001-0.01 Ω cm, thickness: 525 ± 25 μm, 90 nm $SiO_2$, MicroChemicals GmbH, Cat. No. WTD40525255B1011S091)

- 4-inch silicon substrates with 290 nm silicon oxide layer (orientation: <100>, 1-side polished, doping: p-type (Boron), resistivity: 0.001-0.005 Ω cm, thickness: 525 ± 25 μm, 290 nm $SiO_2$, MicroChemicals GmbH, Cat. No. WTD40525205B1050S291)

- 4-inch silicon substrates coated with silicon nitride (orientation: <100>, 2-side polished, doping: p-type (Boron), resistivity: 1-10 Ω cm, thickness: 525 ± 25 μm, 450 ± 50 nm LPCVD $Si_3N_4$ on both sides, MicroChemicals GmbH, Cat. No. WNA40525255B1314S451)

- Hydrogen (99.9995% Research Grade Hydrogen, 200 Bar, BOC Gas & Gear UK, Cat. No. 290626-L)

- Hydrogen in Argon gas mixture (10% Hydrogen / Argon, 200 Bar, BOC Gas & Gear UK, Cat. No. 149776)

- Helium (99.9995%, 200 Bar, BOC Gas & Gear UK, Cat. No. 271088-L)

- Oxygen (99.99%, 200 Bar, BOC Gas & Gear UK, Cat. No. 284916-L)

- Nitrogen (99.9995%, 200 Bar, BOC Gas & Gear UK, Cat. No. 296180-L)

- Argon (99.9995%, 200 Bar, BOC Gas & Gear UK, Cat. No. 293680-L)

- $CHF_3$ (99.99%, 200 Bar, BOC Gas & Gear UK, Cat. No. 153618-L-C)

- $SF_6$ (99.99%, 200 Bar, BOC Gas & Gear UK, Cat. No. 158980-L)

- Ultrapure water with resistivity 18.2 MΩ·cm at 25 °C (Milli-Q® Water Purification System with 0.18 µm filter)

- Methyl isobutyl ketone, MIBK (from Sigma Aldrich, ACS reagent grade, Cat. No. 1061462500) **! CAUTION** This chemical is harmful if swallowed or inhaled. Flammable liquid and vapor. This may form explosive peroxides in air. Can affects central nervous system, liver and kidneys. It can cause irritation to skin, eyes and respiratory tract.

- Sodium hypochlorite solution (Bleach with 6-14% active chlorine, from Sigma Aldrich, Cat. No. 1056142500). **! CAUTION** This is highly harmful & handled carefully with appropriate PPE. Causes severe skin burn and eye damage.

- Ethanol (Fisher Scientific, 99.8%, Cat. No. 12498740) **! CAUTION** This is highly flammable liquid & vapour. There is a serious risk of liquid catching fire; its vapour may catch fire above 13 °C. Keep it away from heat. PPE should be used while handling.

- Acetone (Fisher Scientific, 99.8%, Cat. No. 12498740) **! CAUTION** Highly flammable liquid and vapour. May cause drowsiness or dizziness. Causes serious eye irritation. PPE should be used while handling.

- Isopropyl alcohol (Fisher Scientific, ⩾99.8%, Cat. No. 10533704) **! CAUTION** Highly flammable liquid and vapour. May cause drowsiness or dizziness. Causes serious eye irritation. May cause respiratory irritation. PPE should be used while handling.

- Microposit SI813 G2 Photoresist (Dow Chemical) **! CAUTION** Keep it away from heat, flames, sparks and static discharge. Gloves, safety glass and protective clothing are required when handling. It should be disposed to an approved waste disposal plant.

- Microposit MF 319 Developer (SPEC EM Switzerland) **! CAUTION** The MF 319 developer is hazardous and can cause skin, eye irritation and damage to organs. It should be handled in a special local ventilation. Gloves, safety glass and protective clothing are required when handling. It should be disposed to an approved waste disposal plant.

- Au pellet (99.99 %, Colonial metals), for metal deposition

- Cr pellet (99.99 %, Kurt J. Lesker Company), for metal deposition

- Microposit remover 1165 (Dow Chemical) **! CAUTION** Combustible liquid and vapour. Prolonged, repeated contact with skin may cause drying, defatting, or dermatitis. Incidental contact may cause redness or other transient effects. It is highly toxic and appropriate PPE should be used while handling.

- Polymethyl methacrylate (PMMA, 950K, 3 % wt/wt in Anisole, MicroChem) **! CAUTION** Flammable liquid and vapour. Harmful if inhaled. Causes skin irritation. Causes serious eye irritation. May cause respiratory irritation. Appropriate PPE should be used while handling.

- Potassium hydroxide pellets (Fisher Scientific, Cat. No. 10366240) **! CAUTION** Potassium hydroxide solution is highly corrosive and irritant. May be corrosive to metals. Harmful if swallowed. Causes severe skin burns and eye damage. Harmful if swallowed. It should be handled with proper eye protection and nitrile gloves (standard EN 16523-1:2015). It should be discarded in an appropriate waste disposal container.

- Potassium Chloride (Fisher Scientific, Cat. No. 11459133) **! CAUTION** Maybe harmful if swallowed. Causes serious eye and skin irritation. It should be handled with proper eye protection and gloves.

- Lithium Chloride (Fisher Scientific, Cat. No. 11393258) **! CAUTION** Maybe harmful if swallowed. Causes serious eye and skin irritation. It should be handled with proper eye protection and gloves.

- Aluminum Chloride (Fisher Scientific, Cat. No. 11381649) **! CAUTION** Maybe harmful if swallowed. Causes serious eye and skin irritation. It should be handled with proper eye protection and gloves.

- Sodium Chloride (Fisher Scientific, Cat. No. 11994929) **! CAUTION** Maybe harmful if swallowed. Causes serious eye and skin irritation. It should be handled with proper eye protection and gloves.

- Magnesium Chloride (Fisher Scientific, Cat. No. 10159872) **! CAUTION** Maybe harmful if swallowed. Causes serious eye and skin irritation. It should be handled with proper eye protection and gloves.

- Iron Chloride (Fisher Scientific, Cat. No. 10695862) **! CAUTION** Maybe harmful if swallowed. Causes serious eye and skin irritation. It should be handled with proper eye protection and gloves.

- Indium Chloride (Fisher Scientific, Cat. No. 11384967) **! CAUTION** Maybe harmful if swallowed. Causes serious eye and skin irritation. It should be handled with proper eye protection and gloves.

- Tetramethylammonium chloride (Sigma Aldrich, Cat. No. 74202) **! CAUTION** Maybe harmful if swallowed. Causes serious eye and skin irritation. It should be handled with proper eye protection and gloves.

- Tetraethylammonium chloride (Sigma Aldrich, Cat. No. 86616) **! CAUTION** Maybe harmful if swallowed. Causes serious eye and skin irritation. It should be handled with proper eye protection and gloves.

- Tetrabutylammonium chloride (Sigma Aldrich, Cat. No. 86870) **! CAUTION** Maybe harmful if swallowed. Causes serious eye and skin irritation. It should be handled with proper eye protection and gloves.

- Graphite crystals (GRAPHENIUM, Manchester Nanomaterials) **! CAUTION** Irritating to eyes and respiratory system. PPE should be used while handling.

- $MoS_2$ crystals ($MOS_2$-EL, Manchester Nanomaterials)

- hBN crystals (HBN-A, Manchester Nanomaterials)

- Muscovite mica crystals (MICA-M, Manchester Nanomaterials)

**Equipment and facilities**

- Nitto BT-150E, for graphene, mica and $MoS_2$ (high-tack and high adhesion tape)

- Nitto BT-50E-PR, for large graphene flakes (low residue, high-tack and high adhesion tape)

- Nitto BT-130E- SL (for hBN flake preparation)

- Glass slides

- Cold plate

- Quartz boat

- Molecular sieve desiccant

- Metal plectrum (0.5 mm Stainless Steel, Photofabrication Ltd)

- Micropipettes

- Tweezers with plastic tips

- Tweezers with top fingers and steeped bottom paddle

- Sharp metal tip tweezers

- Silver rod (2 mm in diameter and 500 mm in length, purity: 99.95%, Goodfellow Cambridge Ltd.)

- Faraday cage (length = 30 cm; width = 10 cm; depth: 10 cm)

- Crocodile clips

- Acid resistant O-rings (6 mm and 11 mm in diameter, KALREZ® SPECTRUM DuPont, James Walker UK Ltd)

- A custom-made gas and ion transport cell with screw (for details, see the 'Experimental setup' section of the Introduction)

- Aluminium gravimetry set-up: screws, rubber O-ring

- Gas dosing valve DN 16 with PM-C controller (VAT Vacuum Products Ltd)

- Communication interface (RS-232, Ethernet)

- Keithley 2636B Dual Channel SourceMeter

- AF1 Dual (Elveflow)

- Axopatch 200B (Molecular Devices)
- Helium Leak Detector (Leybold phoenix quadro dry)
- Analog to digital convertors (National Instruments)
- Environmental test chamber (Alpha 335-70 H, Weiss Technik UK Limited)
- Acid and bases wet bench (Clean Air Limited)
- Ultrasonic Cleaner (Fisher Bioblock Scientific, Transsonic TI-H-5)
- Spin-coater (POLOS Spin200i)
- Hotplate (Staurt - SD160)
- Thermal annealing furnace (ThermCraft Incorporate.)
- Electron-beam evaporator (Moorfield, MiniLab 125)
- Optical microscope (Nikon Eclipse LV100ND)
- Transfer stage with optical microscope (Nikon Eclipse LV100ND)
- ProScan$^{TM}$III Motorised Stage System (Prior Scientific)
- Laser writer (Microtech. Model no. LW405, Principle exposure source; 405nm GaN solid-state laser, output 60-100mW)
- Dry etcher (Moorfield Nanoetch using Soft Etching setting and $O_2$/Ar)
- Reactive-ion etcher (Oxford instrument, Plasmalab 100-ICP 65 model)
- Electron beam lithography system (Zeiss SmartSEM + Raith ELPHY Quantum NANOSUITE V6.0)
- Smart SEM (Zeiss)
- AFM Dimension Fastscan with Scanasyst (Bruker)
- Desiccator (Fisher Scientific)
- Microbalance (Mettler Toledo XPE26, with 22g capacity and 1 µg readability)

**Software**
- *LabVIEW* 2018 (National Instruments): https://www.ni.com/en-gb/shop/labview.html
- NIS Elements Viewer: https://www.nikon.com/products/microscope-solutions/support/ and Nikon optical microscope camera: https://www.microscope.healthcare.nikon.com/en_EU/products/cameras/ds-qi2
- Elveflow Smart Interface (ESI) (Elveflow): https://www.elveflow.com/microfluidic-products/microfluidics-software/
- OriginPro® (OriginLab): https://www.originlab.com/index.aspx?go=Products/Origin
- Clewin Software: https://wieweb.com/site/product-category/clewin-software/
- Inkscape: https://inkscape.org/release/inkscape-1.2.2/

**Procedure**

**Substrate Fabrication     Timing 1-3 d**

**CRITICAL STEP:** All these steps need to be carried out in ISO class 5 and 6 cleanrooms. The substrate fabrication process is summarized in Fig. 5.

1. Design the photolithography patterns in Clewin software, to be exposed with laser writer. Window dimensions (800 µm × 800 µm) in the design were chosen considering the desired $SiN_x$ membrane dimensions (~100 µm × 100 µm) after KOH etching. The windows were spaced on a 4-inch wafer in a way that individual devices can be easily separated by cleaving the pieces using a wafer scriber.

**CRITICAL STEP:** Here we chose the distance between the two windows to be 17,000 µm and therefore could make 26 windows on a 4-inch wafer. If the distance between the two windows is decreased, we can increase the number of windows on the size of wafer. For example, on reducing the distance between the windows to 11,000 µm we can get 49 windows (Fig. 15).

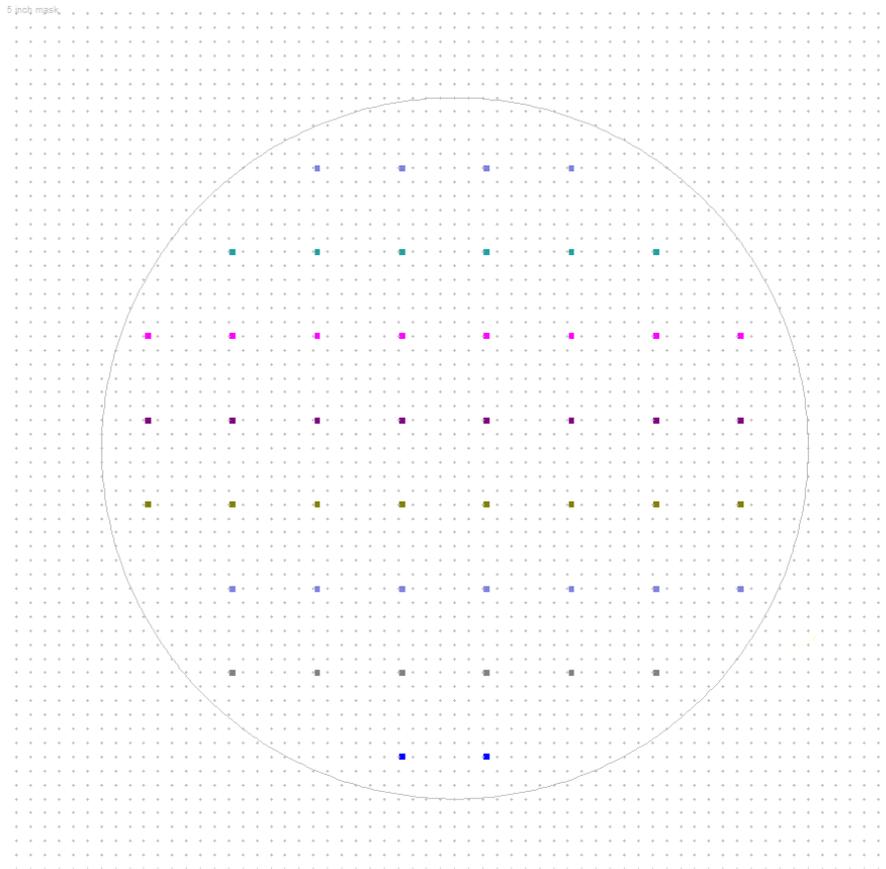

**Fig. 15 |** Clewin design file for 4-inch wafer, with a distance between two windows is 11,000 µm. Each row of windows is kept in a different layer in the software (hence seen in different color) to ensure faster patterning with laser writer.

2. Clean a 4-inch double side polished silicon nitride (500 nm on both sides) wafer with acetone and IPA followed by drying it under flow of nitrogen using a nitrogen gun.

3. Spin coat one side of the silicon nitride wafer with S1813 positive photoresist using the recipe R1 (Table 1).

**! CAUTION:** The photoresist layer should be spread uniformly without any bubbles before spinning.

**CRITICAL STEP:** Use tweezers with top fingers and steeped bottom paddle to handle the full wafers and use a 4-inch wafer on vacuum chuck of the spinner to prevent any physical damage. Clean all the surfaces (spin coater, hot plate and cold plate) using IPA before using to prevent any contamination with the wafer.

4. Spin-coat the other side of the wafer with S1813 positive photoresist using the recipe R1 (Table 1). Soft baking is done to drive off some of the residual solvent in the photoresist and partially solidify.

5. The wafer is exposed by photolithography. We use a laser writer (wavelength: 405 nm, lens: 3, Filter: 10%, Energy: 387 mJ/cm$^2$) and the desired pattern is exposed using photon steer software.

6. Post exposure, develop with MF-319 developer for 60 s, rinse with DI water for 30 s, and dry it with a nitrogen gun.

7. Inspect the wafer under optical microscope using red or yellow filter to check if the windows have developed.

**! CAUTION:** Use of a red filter is mandatory otherwise white light from microscope will further expose the resist. Green patches within the exposed areas indicate insufficient developing time or underexposure of laser. If this is the case, clean the wafer in hot acetone (~ 50 °C) for 10 minutes and then in IPA for 10 minutes to remove photoresist and restart the photolithography process.

**CRITICAL STEP:** If the wafer is still not clean and photoresist patches are clearly visible then leave the wafer in a solution of MF-319 overnight and then clean with microposit 1165, then with IPA and then dry using nitrogen gun.

8. PlasmaLab 100-ICP 65 model RIE is used to etch the SiN$_x$ layer with the recipe R2 (Table 1).

**! CAUTION:** Always clean the etching chamber with oxygen plasma before and after using RIE to avoid contamination.

**CRITICAL STEP:** Avoid continuous etching over 3 mins as this may cause the photoresist to cross-link and resulting in unclean substrates.

9. Clean the wafer with hot acetone (50 °C) and IPA for 5 minute each to remove the photoresist and then dry using nitrogen gun.

**CRITICAL STEP:** If the resist is not removed, use MF-319 developer or microposit -1165 solution for overnight immersion of wafer to remove the photoresist.

10. Keep the wafer in 30% w/v aqueous KOH solution with a volume of 1.5 L at 80 °C for etching the Si layer.

11. After hot KOH etching for 8 hours, the wafer is washed with ample amounts of DI water and inspected for the formed SiN$_x$ membrane windows. If left for a longer time, the size of the membrane may increase due to etching in <100> direction. Repeat the hot KOH etch for another 30 mins if the membranes are not formed.

**! CAUTION:** The etching time can slightly vary for different membrane sizes.

12. The SiN$_x$ wafer is washed with DI water and IPA after hot KOH etching and then dried under the flow of N$_2$.

13. Check the wafer under microscope for the cleanliness of SiN$_x$ membranes in the wafer with help of both the front and back light illumination.

**Second photolithography to make a hole in the SiN$_x$ membrane**

14. Scribe the full wafer substrate (this should be 11 mm × 11 mm) with a SiN$_x$ membrane (100 µm × 100 µm) such that the membrane is exactly in the centre of the substrate.

15. Spin-coat S1813 resist on the substrate. To prevent the membrane from breaking by the spin-coater vacuum, stick the high-tack tape on the backside (not on the membrane side) of the substrate.

16. Remove the high-tack tape and bake the substrate with the recipe R1 (Table 1).

17. To make the rectangular hole of 3 µm × 25 µm at the centre of SiN$_x$ membrane (100 µm × 100 µm), locate the membrane area of interest using lens 3 of the laser writer.

18. Find the centre of the membrane and assign it as origin (0,0) on the laser writer.

19. The membrane is exposed in the laser writer (wavelength: 405 nm, lens: 5, Filter: 10%, Energy: 387 mJ/cm$^2$) to write the rectangular hole 3 µm × 25 µm pattern which is designed using Clewin software.

**CRITICAL STEP:** Decrease the lens number (to 5) and use the same energy as used above in step 5. The energy/dose depend on the instrument model/make and it may vary significantly. This dose is optimized for the in-house laser writer in our clean room.

20. After writing the rectangular hole pattern, develop the substrate in MF319 solution for 60 sec and 30 sec in DI water and then dried using nitrogen air gun.

21. Developed pattern is checked under microscope using red/yellow filter to find if it is under or over developed similar to step 7.

22. Dry etching is performed on the substrate as described in step 8 to etch the SiN$_x$ layer with the recipe R2 (Table 1) and further resist cleaning is performed. The optical image of the substrate with ~ 3 µm × 25 µm hole on SiN$_x$ membrane is shown in Fig. 5h.

**Mechanical exfoliation and flake preparation    Timing 1-2 d**

**CRITICAL STEP:** Mechanical exfoliation and flake preparation is performed in a clean room facility (class 100).

23. Clean a SiO$_2$/Si chip (20 × 24 mm scribed piece from 4-inch wafer with 290 nm thick silicon oxide layers, single side polished) in acetone and IPA on ultrasonic bath with the recipe R3 (Table 1), followed by drying with nitrogen air gun. Multiple chips can be cleaned at once to increase the substrate yield.

**! CAUTION:** This step is to clean the chips to remove wafer scriber particles, resist residues and chemicals adsorbed on surfaces.

24. After wet cleaning, the SiO$_2$/Si chips are dry cleaned using recipe R4 in nanoetcher (Table 1). Oxygen plasma cleaning is carried out to remove any contaminants and increase adhesion of flakes to substrates.

25. Natural graphite crystal (bulk crystal with shiny facets in the sizes of few millimeters to a centimeter) is pressed onto a 1-inch square piece of high-tack tape (Nitto 150E - high-tack and high adhesion tape). This tape is then removed from the graphite crystal and peeled repeatedly on new tapes to produce thin graphite flakes on the tape.

**TROUBLESHOOTING**

26. Thin graphite flakes on the adhesive tape are then pressed on a SiO$_2$/Si substrate just after the substrate is removed from oxygen plasma etcher.

**TROUBLESHOOTING**

27. The substrate with tape is kept for 4-5 hours and then the high-tack tape is peeled off with a low angle to remove thick graphite and leave only thin graphene behind on the chips.

**CRITICAL STEP:** A similar step is repeated to place other thin 2D crystal flakes on SiO$_2$/Si substrates such as hBN, MoS$_2$ and mica. Nitto BT-150E tape is used to exfoliate mica and MoS$_2$ on 290 nm SiO$_2$/Si substrate while Nitto BT-130E- SL tape is used to exfoliate hBN on 90 nm SiO$_2$/Si substrate.

**CRITICAL STEP:** Optical inspection is critical and require enough experience in identification skills to distinguish between different thicknesses of 2D crystal flakes on SiO$_2$/Si substrate.

28. A very quick way of optical identification is to generate the intensity plot for the flake of 2D materials on SiO$_2$/Si substrate. Intensity line plot eases out the ambiguity between numbers of layers that's been exfoliated on the substrate.

29. Graphene on 290 nm SiO$_2$/Si surface can be easily observed using optical microscope without any color filter, differential interference contrast (DIC) mode or in dark field mode. (Fig. 6a). On 290 nm SiO$_2$, pale violet can be bilayer to 5 layers, dark violet indicates 3-5 nm (~10-15 layers), and yellow color indicates thickness >50 nm.

**CRITICAL STEP:** About 10 nm to 15 nm thick flakes with uniform layer (uniform color, without steps on flakes) are chosen as bottom layers for the 2D channel devices. It will be difficult to etch using the optimized etch recipe if thicker flakes (> 30 nm) are chosen for bottom crystals.

30. A thin crystal of graphite, hBN or MoS$_2$ (~10-15 nm in thickness) is used as a bottom layer for the devices, while thick crystal of graphite, hBN or MoS$_2$ (~70-150 nm in thickness) are used as a top layer for the devices. (See Fig.6 for optical images of crystals used for top and bottom)

**Spacer fabrication      Timing 2-3 d**

31. Spacer flakes are prepared by following steps involved in 'flake preparation'.

32. Monolayer or few layer graphene flakes are chosen for further processing as detailed below to prepare parallel strips graphene, which will act as spacer in our 2D channel devices.

33. Spin-coat a positive e-beam resist PMMA on the substrate containing chosen flake using the recipe R6 (Table 1).

**CRITICAL STEP:** Clean the substrate's backside surface as some of the resist (PMMA) might stick or overflow on the edges/sides during spin coating. This would lead to uneven surface on the backside of the substrate and result under/over exposure in the EBL step due to non-uniform e-beam focus on the PMMA resist.

**! CAUTION:** Handle the substrates only with non-scratching tweezers to avoid scratching the thin films.

34. Take the stage coordinates to define EBL global coordinates for writing alignment marks around the chosen flakes.

35. Place the wafer onto a microscope stage and define a global origin point (south west, SW) and angle correction point (south east, SE) as global coordinates for the substrate (illustrated in Fig.7a).

The angle between two marks (SW and SE) should be around 180±2°. Take the flake coordinates with reference to origin point (SW) and angle correction.

**TROUBLESHOOTING**

**CRITICAL STEP:** While taking the stage coordinates, make sure that sample does not move after the angle correction. It will change the flake position with respect to reference points (SW and SE) and lead to exposure of alignment marks at a wrong place after EBL.

**PAUSE POINT:** The substrate can be stored in a clean and dust-free environment.

36. Expose the flake for alignment marks in EBL. We use 1000 µm × 1000 µm writing field and write cross marks (as showed in Fig.7) spaced by 200 µm to each other around the chosen flake.

**PAUSE POINT:** The substrate can be stored in a clean and dust-free environment.

37. After exposure, develop the markers (alignment marks) in 1:3 vol/vol MIBK/IPA mixture at room temperature or 1:3 vol/vol water/IPA mixture at 5 °C for 45 seconds followed by immersing in IPA for 30 seconds to stop further development, and then blow dry with nitrogen.

38. Take the images of alignment marks with optical microscope at different magnification i.e., 5×, 10×, 20× and 50×. Scale bar should be calibrated and correctly labelled for every image.

**PAUSE POINT:** The substrate can be stored in a clean and dust-free environment.

**CRITICAL STEP:** To define the trenches/lines for nanochannels, make the patterns using Inkscape software and convert the design file in GDSII format using Clewin software.

39. First, embed all the images of different magnifications in an Inkspace file and ensure that all images of different magnification are overlapped/superimposed.

40. For designing the pattern on the flake, make several hundred (typically 200, depends on the size of the flake) lines of 100 nm width separated by 170 nm in the Inkscape and save the file in DXF (or better DXF) format.

41. Using Clewin software, open DXF (or better DXF) file, select the designed pattern (select all), and connect all the lines using the "Connect wires" from 'Edit' option. Make sure the file is saved as GDSII which is a readable file format on the Raith (beam blanker/controller) software in the EBL system.

42. Expose the patterned channels design in EBL at 20 kV using an area dose of 210 µC/cm$^2$ to create 100 nm wide nanochannels in the PMMA resist. Use patterned marks to locally align the nanochannel patterns to the flake as shown in Fig 7.

43. Develop the exposed lines in 1:3 vol/vol MIBK/IPA mixture at room temperature or 1:3 vol/vol water/IPA mixture at 5 °C for 30 seconds followed by immersing in IPA for 30 seconds to stop further development, and then blow dry with nitrogen.

**TROUBLESHOOTING**

44. Using an optical microscope (under 50× or 100× magnification) check if the pattern is developed on the flake.

    Inspect the pattern by AFM imaging to ensure that the lines are exposed properly and deep enough through the resist. With the above-mentioned EBL resist spin-coat recipe will result in the thickness of PMMA between 130 to 150 nm. In ideal condition, the channel/spacer line width and depth of PMMA attain dimensions of (100/160±20) nm and 120±20 nm

respectively. Figure 8 shows the AFM images of lines with their height profile and different conditions that occur after exposure and development.

**TROUBLESHOOTING**

**CRITICAL STEP:** Always make sure to develop for 30 (± 2) sec and not more than that otherwise it will lead to overexposed lines rendering very narrow spacers after etching.

**PAUSE POINT:** The substrate can be stored in a clean and dust-free environment.

45. Etch the PMMA patterned flake using moorfield nanoetcher with the recipe R5 (Table 1). The total etching time depends upon the thickness of the spacer flake. Usually, 20 sec is required for etching one layer of graphene using our nanoetcher.

**TROUBLESHOOTING**

**CRITICAL STEP:** Do not etch for more than 30 sec in one go, rather break the total etching time into small step of 20 sec with the gap of 5-6 sec between etching steps. This is important to avoid cross linking of the resist (PMMA) otherwise it will be difficult to remove afterwards in cleaning step.

46. After etching the exposed graphene using PMMA as mask on nanoetcher, remove the resist in acetone with mild sonication recipe R7 (Table 1).

**CRITICAL STEP:** Do not put the substrate directly into acetone beaker placed in the sonicator bath. Hold the substrate with the tweezer and slowly take it in and out of the acetone for every 5 sec. Take the substrate out of solution only for a short time period (~1 to 2 sec) to avoid drying of acetone. Dry acetone leaves behind residue marks, which are difficult to remove afterwards.

47. Redoing hot acetone treatment at 50 °C for few hours (2 – 8 hrs until clean) followed by IPA wash and blow dry using nitrogen gun.

**TROUBLESHOOTING**

48. Check the cleanliness of the spacers under AFM and if they are not clean repeat above steps 46 and 47.

49. A thermal annealing process at 400 °C in inert atmosphere (10 % $H_2$ in Ar) for 4 hours can be carried out to remove very thin residues of PMMA resist on spacer lines.

**CRITICAL STEP:** The annealing should take place after the spacers have been cleaned thoroughly in acetone, otherwise it may cause PMMA to harden, and cleaning may become more difficult. After every cleaning procedure, nanochannels are scanned under AFM to check the cleanliness and quality of the spacers.

**Transfer of flake          Timing 2-3 d**

50. Transfer the chosen bottom layer (graphite, hBN or $MoS_2$) on to the rectangular hole in the $SiN_x$ membrane using wet transfer methods and PMMA as polymer support.

**CRITICAL STEP:** By using a small triangular strip of adhesive high-tack tape, an area within a few mm around the chosen flake should be cleared of thick 2D crystals. It can be cleaned by sticking the tape on the thick crystals with tweezers while not going near the selected bottom flake.

51. Spin-coat the substrate having bottom flake with PMMA with the recipe R6 (Table 1).

52. A tape window is prepared by cutting a piece of Nitto 150E (high-tack tape) and making 1 mm$^2$ hole in the centre of the tape by using a punching tool.

**CRITICAL STEP:** The circular window should be cleaned off any sharp edges which may potentially break the PMMA membrane at later stage. For this, we can use a sharp tweezer to pluck the sharp protruding pieces inside the circular hole.

53. Place the tape window around the flake, so that the flake is at the centre of the hole. Make sure the tape (adhesive side) is completely in contact with PMMA resist without any air bubbles or gaps between them.

**TROUBLESHOOTING**

54. Scratch around the tape using a sharp metal tip to disconnect PMMA under the tape from the substrate for lifting.

55. Prepare 3% (wt/vol) KOH aqueous solution in a PTFE bowl for etching SiO$_2$ layer (sacrificial layer) on the chip, to lift the PMMA membrane with flake, which are stuck to tape window.

56. Leave the substrate in aqueous KOH solution for 5 to 8 hours at room temperature.

**CRITICAL STEP:** This will etch the SiO$_2$ and make the tape window attached to PMMA to lift off on the surface of solution.

57. The PMMA tape window is dipped in DI water to remove the KOH residue and kept in a beaker full of fresh DI water.

**PAUSE POINT:** The PMMA tape window can be kept in water for up to 24h in a clean room environment before proceeding for the next step.

**CRITICAL STEP:** This method can be used to lift bottom, spacer or top crystal of any 2D materials such as graphite, hBN, mica or MoS$_2$, for the fabrication of Å-channels. The bottom dimensions (width and length) should be at least 50 µm × 60 µm for devices made on 3 µm × 25 µm SiN$_x$ hole, and thickness of bottom flake should be between 10 to 20 nm. Top crystals should be 25 µm × 50 µm in dimensions (width and length) and thickness should be at least 70 nm to 150 nm to provide enough mechanical rigidity to avoid sagging in to the channels.

58. PMMA tape window with the flake is taken from the DI water and placed on a metal plectrum with the aid of double-sided tape. The plectrum is then loaded on the vacuum chuck in such a way that the flake would face downward (Fig.10).

**CRITICAL STEP:** This step is carried out carefully, as the PMMA membrane might break if not handled properly.

**TROUBLESHOOTING**

59. Using a high-tack tape, cover the bottom side of SiN$_x$ membrane substrate to avoid cracks or damage to freestanding SiN$_x$ membrane due to the vacuum holder of the transfer stage.

60. Place the substrate on the vacuum chuck of the transfer stage and switch it 'ON'. Set the temperature of the stage (substrate holder) to 65 °C.

61. Place the plectrum with PMMA membrane on the plectrum holder arm after switching 'ON' plectrum vacuum to hold it firmly.

62. Using the optical microscope, annotate both flake on the PMMA membrane and rectangular hole on the substrate. Rotate/move the stage to align the hole position to the flake position first with 5× magnification and then at higher magnifications, until 50×.

63. Move the plectrum arm down (along z- axis) towards the substrate until it is near the substrate (< 1 mm).

64. Align the flake and desired location on the substrate (rectangular hole) by moving the plectrum arm in x- or y- direction.

65. Make sure the flake and the hole are aligned well by carefully moving optical microscope focus between flake on the PMMA membrane and hole on the substrate.

66. Continue to lower the plectrum arm toward the substrate. Adjust the focus, and then move the arm down with z movement.

67. Repeat the above step until the flake and hole on the $SiN_x$ membrane are both in focus at the same time, or the plectrum arm's z- movement provide resistance while further lowering.

68. At 5×, check whether the flake is completely in the hole as per the initial alignment.

**TROUBLESHOOTING**

69. Make sure all the part of PMMA membrane has touched the substrate and slowly settles on the substrate's surface without air gaps or folds. If not, use blunt metal tweezer to touch it.

70. Scratch off the PMMA near the edges to separate the region of interest from the tape with hole. Move the holder up, leaving behind the PMMA membrane with bottom flake on $SiN_x$ hole.

71. Switch off the temperature and the vacuum and unmount the substrate. Remove the bottom high-tack tape before further processing the substrate.

72. Bake the substrate with bottom flake and PMMA at 140 °C for up to 15 minutes to increase the adhesion or contact between flake and the substrate well.

73. Stick high-tack tape again on the back side of the substrate before removing the PMMA by placing it in beaker filled with acetone for 15 minutes at room temperature.

74. Clean the substrate with acetone, IPA and dry using nitrogen gun.

75. Check the sample under optical microscope if there are any residues of polymer or air bubbles under the bottom flake.

76. If there are any residues of PMMA or tape residues, cover the backside of the substrate with tape and leave it in the hot acetone (~50 °C) for 30 minutes. And repeat above steps 74 and 75.

77. After the transfer of bottom on the rectangular hole, bottom flake is back etched using RIE to replicate the hole into the bottom flake. Use recipe R8-Table 1 for graphite bottom and R9-Table 1 for hBN bottom.

78. After preparing spacers on mono- or few layer 2D crystals, wet transfer process (similar to steps 50-77) of bottom and spacer is performed.

**CRITICAL STEP:** Make sure proper alignment is done before transferring spacer on to the bottom flake. Spacer lines should be perpendicular to the rectangular hole (long axis) in the bottom flake.

79. After the transfer of spacers on to the bottom, they are etched from the back of the substrate using RIE. Use recipe R8-Table 1 for graphite bottom and R9-Table 1 for hBN bottom.

80. Wet transfer method (step 50-77) is carried out for placing the top crystal on the stack of bottom and spacers. One can also use any of the dry transfer methods described in this protocol for the top layer transfer.

**Gold patch and Device post-processing       Timing 1-2 d**

81. After successful transfer of tri-crystal stack of bottom, spacers and top crystals, we do further lithography processing for defining channel length. Before and after the post-processing steps, the Å-channels devices are annealed in a thermal annealing furnace under a gas mixture of 10% $H_2$ and 90% Ar flowing at a rate of 0.15 L/min.

**! CAUTION:** Device with $MoS_2$ crystal should be annealed at 300 °C for 3 hours and 350 °C for 3 hours. Likewise, device with graphite or hBN crystals should be annealed at 300 °C for 3 hours and 400 °C for 4 hours.

82. Take the optical image of the device under the optical microscope with the objective of 5×, 20×, 50× and 100× respectively. Scale bar should be calibrated before the use and should be correctly labelled for every image.

83. Use Inkspace software to embed the 50× optical image and ensure the scale of the drawing to be in line with the scale bar in the embedded image.

84. Make a rectangular patch design where the strip should be aligned along the hole of the $SiN_x$ membrane. Transfer the design profile into the .cif format.

**CRITICAL STEP:** The patch design will be used as etching mask to manipulate the channel length. One should make the design carefully as it determines the location and number of channel openings.

85. Attach a high-tack tape on the back side of device to prevent the contamination from chemicals entering micro cavity and to provide device with a buffer when confronting the vacuum force.

**CRITICAL STEP:** Back side of device refers to the side with the big hole on the Si substrate while the front side is the tri-crystal stack side.

86. Write the gold patch design either using photolithography or electron beam lithography on the bottom-spacers-top stack of the device coving the rectangular hole on $SiN_x$ membrane.

**Photolithography to write the gold patch design     Timing 3 hours**

**CRITICAL STEP:** This section (steps 87-92) can be skipped if using the electron beam lithography for the pattern writing.

87. Spin coat the S1813 photoresist onto the device with the recipe R1 (Table 1).

88. Remove the tape from the back side of the device and bake the freshly coated device at 110 °C for 60 s.

89. Expose the pattern using the laser writer with a laser wavelength of 405 nm, lens 5, 10% filter and a full gain.

**CRITICAL STEP:** Pattern design should be accurately positioned and exposed.

90. Place a low adhesive tape to the back side of device and develop the pattern by immersing the substrate into MF 319 developer for 45 second, followed by another 45 second in the DI water.

91. Dry the substrate under flow of nitrogen and check the developed pattern under the optical microscope with a red/yellow filter of 450 nm or more than that.

92. Remove the tape from the back side of the device. The device is ready for the subsequent metal deposition procedure.

**PAUSE POINT:** The device can be stored in a clean and dust-free environment.

**Electron beam lithography to write the gold patch design    Timing 4 hours**

**CRITICAL STEP:** This section (steps 93-107) can be skipped if using the photolithography for the pattern writing.

93. Spin coat a PMMA solution on the device using the recipe R6 (Table 1).
94. Remove the tape from the back side of the device and bake the PMMA coated device at 150 °C for 5 minutes.
95. Expose the pattern using the electron beam with either 10 kV or 20 kV, aperture 20 µm, spot size 250, dosage 300 µC/cm$^2$.

**CRITICAL STEP:** Pattern design should be accurately positioned and exposed.

96. Place a low adhesive tape to the back side of device and develop the pattern by immersing the substrate into a cooled developer DI: IPA water (1/3, vol/vol) for 45 second, followed by another 45 seconds in the pure IPA solution.
97. Dry the substrate using nitrogen air gun and check the developed pattern under the optical microscope.
98. Remove the tape from the back side of the device. The device is ready for the subsequent metal deposition procedure.

**PAUSE POINT:** The device can be stored in a clean and dust-free environment.

99. Onto the substrate with patch design patterned by lithography, deposit a metal layer of Cr (5 nm) and Au (50 nm).
100. Use a low adhesive tape to cover the back side of the substrate when doing the metal lift-off. Soak the metal deposited device into a hot acetone (~ 50 °C) for 2 hours.

**CRITICAL STEP:** Check if metal film on the device surface has formed any wrinkles or swelling. If that happens, carefully remove hot acetone beaker from hot plate and let it cool down to clean metal film on the substrate.

101. Use a clean pipette to blow the acetone onto the substrate surface immersed in the acetone solution to do the metal lift-off process.

**CRITICAL STEP:** The blowing pipette should be at least at ~ 0.5 to 1 centimetre above the substrate surface in acetone.

102. Transfer the device to another fresh hot acetone solution (50 °C) for 15 minutes.
103. Place the device into an IPA filled petri dish. Closely check the metal lift-off condition (i.e., continuous Au strip with clear boundaries, no overhanging pieces of Au layers) under the optical microscope with different objectives.
104. Repeat step 102-103 until a clear gold design patch can be seen under optical microscope.

**CRITICAL STEP:** Device should be always in a wet condition before the completion of metal lift-off procedure to avoid dried acetone residues.

105. Remove the attached tape from the back side of the device. Clean the device with acetone, IPA and dry using the nitrogen gun.

**PAUSE POINT:** The device can be stored in a clean and dust-free environment.

106. Etch the device using RIE. For bottom graphite-graphene spacer-top graphite device, etch the device with recipe R8 (Table 1). For bottom hBN-graphene spacer-top hBN device, etch the top hBN layer with recipe R9 (Table 1). This recipe gives an etching rate of 5 nm hBN per second.

107. Device is imaged under optical microscope and images of final device are recorded in all magnifications and illumination modes.

**PAUSE POINT:** The device can be stored in a clean and dust-free environment.

**Gas transport measurements        Timing 1-2 d**

108. Mount the fabricated device in device holder and screw tight it. Device holder is placed in the helium leak detector setup as shown in Fig 12.

**CRITICAL STEP:** When loading the device into the device holder make sure the two O-rings are well-aligned with each other. If the O-rings are not well aligned, the device may break/crack while loading the device holder into the helium leak detector.

**TROUBLESHOOTING**

109. Evacuate the helium leak detector setup down to 1 mbar.

110. Fill the He gas in the balloon from the gas cylinder and close the neck of the balloon with stopper to avoid leakage of the gas.

111. Load the gas balloon with the stopper on to voltage operated dosing valve which is connected to the helium leak detector assembly. Evacuate the setup including device chambers and loop lines down to 1 mbar.

**CRITICAL STEP:** this step is important as it will evacuate any traces of the gas from the balloon opening and to avoid mixing of gases from atmosphere.

112. Stop the dosing valve and remove the stopper on the helium balloon.

113. Turn the leak detector 'On' and select the mass 4 for He leak detection and wait for it to get stabilized down to $\sim 10^{-12}$ mbar litre/sec detection values on the display.

114. Open the leak detector valve attached to the device setup assembly and wait for the leak detector value to reach the value between $10^{-10}$ and $10^{-12}$ mbar litre/sec.

115. Start the *LabVIEW* recording program to collect the leak rate and pressure data.

116. Set a gas flow rate of the dosing valve using its controller to purge the gas into the channels and record it till the pressure in the feed chamber reach to 500 mbar.

117. After each measurement, evacuate the setup again down to 1 mbar and take another measurement. Collect five sets of data on each device.

118. Process and analyse the acquired data as shown in the section "Data analyses-gas transport".

**Water transport measurement        Timing 3-5 d**

**CRITICAL STEP:** Prior to the water transport measurement, test the device with the helium leak detector to ensure the channels are open.

119. Micro balance is placed in the environmental chamber which maintains required temperature (in the range of 20 °C to 40 °C) and relative humidity (RH from 20 % to 90 %). the microbalance is connected to computer through *LabVIEW* program to collect data.

120. Use the DI water to partially fill (0.5 ml) the steel container, gravimetric cell as shown in Fig. 13).

121. Mount the device into the gravimetric cell set up ensuring the device is securely placed between O-rings of container cell and lid. Then it is carefully screwed to tightly close the cell (Fig. 13).

**! CAUTION:** Care must be taken while aligning screws between lid and container. If a screw touches the device edge, it will lead to breaking of the $SiN_x$/Si substrate. Screws should be fastened gently by tightening opposite positions and avoiding the application of unequal pressure/force on the device. The device substrate and O-rings should be clean and free of any dust particles to avoid unwanted leaks.

122. Place the cell into the microbalance. Wait for the system to stabilize.

123. Start the *LabVIEW* program to record the weight of the microbalance with an interval time of 1 minute along with temperature and relative humidity from the environmental chamber.

124. Monitor the measurement for several days and collect data.

125. Process and analyse the acquired data, as shown in section "Data analyses-water transport".

**Ionic curren*t* (I-V) measurement Timing 1-3 d**

**CRITICAL STEP:** Place the device in the device holder as displayed in Fig. 14 with the help of acid resistant O-rings and tighten the screws carefully.

126. Wash the device by filling the reservoirs with the following order of liquids. First start with DI water (3-5 times) then with mixture of IPA and DI water in 1:3 ratio (2 times) and finally with DI water (3 times).

127. Fill the reservoirs (approximately 1.5 ml to 2 ml) with the desired concentration of electrolyte solution.

**CRITICAL STEP:** To remove any trapped air bubbles inside the reservoir, do repetitive rinsing with electrolyte using micropipette.

128. Insert the Ag/AgCl electrodes in both reservoirs as shown in Fig. 14.

**CRITICAL STEP:** Prior to the measurements, always clean the electrode with ethanol followed by DI water.

129. First connect the negative terminal of the Keithley (Low) to the electrode placed in channels side reservoir and later connect positive terminal of Keithley (High) to electrode placed another reservoir.

130. Fill both the reservoirs with the same concentration electrolyte solution. For the first test, use the lowest concentration that will be tested (for example, $10^{-6}$ M electrolyte solutions in our case).

131. Both the reservoirs need to be filled with the same electrolyte solution (same concentration) and similar volume, ~ 2 ml.

132. Measure the current through the device while varying the voltage between ±200 mV. Repeat for 3 loops while recording measurement data with *LabVIEW* program.

133. Steps 126-129 are repeated for other concentrations of electrolytes, for example from $10^{-5}$ M to 1 M and so on.

134. Measured *I-V* values are analysed as in section "Data analyses – ion transport".

    Fill the reservoirs with equal volume (~2 ml) of desired electrolyte solution, but with a different concentration in each side. For example, $10^{-1}$ M KCl solution in the first and $10^{-2}$ M KCl solution in the second reservoir.

**CRITICAL STEP:** Use the similar device configuration as depicted in the setup Fig. 14.

135. Insert electrodes in the cell and first connect the 'Low' (negative) terminal to channel side electrode and 'High' (positive) terminal to another reservoir electrode.

136. Perform current versus time measurements to ensure saturation of current. This will typically take 10 to 20 min for the current to stabilise.

137. After saturation, immediately start the *I-V* measurements with voltage range ±200 mV for 3 loops.

138. Remove the Ag/AgCl electrodes and insert the commercially available (saturated salt bridge) Ag/AgCl reference electrodes.

139. Repeat the *I-V* measurements with the same voltage range ±200 mV, for 3 loops with the reference electrodes.

**Pressure cell mounting for streaming        Timing 3-6 hours**

140. Mount the device in the pressure cell and wash several times (following the device washing procedure, Step 126). Fill the reservoir with the desired electrolyte.

141. Then place electrodes which are pre-inserted into pressure cell caps and hand tight caps gently.

**CRITICAL STEP:** The pressure cell should be leak-tight. If any leakage occurs, stop the procedure, and check the seal.

142. Place the cell into a faraday cage and connect the electrodes (see step 129) with the source meter.

    Note: In the terminology used in this protocol, if pressure is applied to the channel side it is called negative pressure and if applied to the hole side it is called as positive pressure.

143. Connect pressure 'inlet' pipe to cell & close cage as shown in the Fig.14c.

144. Start the programme in ESI Elveflow software.

Table 1: Process recipes and parameter table

| Recipe code | Type of recipe | Instrument used | Recipe details | Post processing and other information |
|---|---|---|---|---|
| R1 | Spin-coating photoresist (S1813 or S1805) | Spin-coater (POLOS Spin200i) | Speed: 3000 rpm. Acceleration: 1000 rpm/s. Time: 60 s. | Soft bake for 60 s at 110 °C and then on cold plate for 15 s. |
| R2 | Dry etching of $SiN_x$ (Reactive ion etching) | Plasmalab 100 ICP-65 (Oxford) | Gases used: $CHF_3$ and $SF_6$ in the ratio 60:15. Pressure: 3mTorr RF power: 30 W | Used to remove $SiN_x$ during the fabrication of $SiN_x$ membranes and holes. |

| | | | ICP: 600 W<br>Temperature: 10 °C<br>Etch cycle duration: 30 sec.<br>6 cycles for 500 nm thick $SiN_x$ with total etching of 3 mins.<br>2 cycles for 100 nm thick $SiN_x$ with total etching of 1 mins. | |
|---|---|---|---|---|
| R3 | Substrate Cleaning (wet) | Ultrasonicator (Fisher Bioblock Scientific, Transsonic TI-H-5) | Power: 100%<br>Frequency: 130 kHz<br>Mode: normal<br>Time: 10 min | Solvent used: acetone followed by IPA. Acetone to remove polymer and organic contaminants. IPA to clean residues & avoid solvent marks. |
| R4 | Substrate Cleaning (dry) | Nanoetcher (Moorfield) | RF PSU: 12 W<br>Gases used: $O_2$ and Ar in the ratio of 16:8.<br>Pressure: $26 \times 10^{-3}$ mbar<br>Time: 10 min | Remove hydrocarbon contamination and increase surface adhesion. |
| R5 | Graphene spacers etching | Nanoetcher (Moorfield) | RF PSU: 12 W<br>Gases used: $O_2$ and Ar in the ratio of 16:8.<br>Pressure: $26 \times 10^{-3}$ mbar<br>Time: ~20 s per layer | Repeated cycles of etching with 20 s exposure of plasma for multi-layer graphene spacers. For example, for tri-layer graphene spacers, three cycles of 20 s etching with a pause of 5 s in between. |
| R6 | Spin-coating (PMMA 950K 3% in anisole) | Spin-coater (POLOS Spin200i) | Speed: 3000 rpm.<br>Acceleration: 1000 rpm/s.<br>Time: 60 s. | Soft bake for 5 min at 150 °C and then on cold plate for 15 s. |
| R7 | Mild sonication for cleaning PMMA | Ultrasonicator (Fisher Bioblock Scientific, Transsonic TI-H-5) | Power: 10%<br>Frequency: 135 kHz<br>Mode: degas<br>Time: 60 s. | Mild sonication for cleaning spacers. Lift the substrate from the acetone for every 10 seconds to avoid damage to spacer flake. |
| R8 | Dry etching (Graphite) | Plasmalab 100 ICP-65 (Oxford) | Gas used: $O_2$<br>Pressure: 3 mTorr<br>ICP 600 W<br>RF Power 30 W<br>Temperature 10 °C<br>Time: 3 – 5 min | 5 minutes etching for bottom graphite in graphite-graphene spacer-top graphite devices after gold patch.<br>3 minutes etching for bottom graphite from backside of the substrate. |
| R9 | Dry etching (hBN) | Plasmalab 100 ICP-65 (Oxford) | Gas used: $CHF_3$ and $O_2$ in the ratio 60:15.<br>Pressure: 10 mTorr<br>ICP 300 W<br>RF Power 5 W<br>Temperature 10 °C<br>Etch rate: 3 nm/s | For etching hBN in bottom hBN-graphene spacer-top hBN devices after gold patch. |

Table 2: Troubleshooting advice.

| Step | Problem | Possible reason | Solution |
|------|---------|-----------------|----------|
| 25 | Small white color particles or contamination on bulk graphite crystals | Random shapes | Cleaning with sticky tape and peel-off contaminated areas. Use relatively flat surface of the bulk crystal for flake preparation. |
| 26 | Breaking of chip | Uneven pressing | Apply pressure evenly on the substrates while flake preparation. |
| 35 | Error in stage coordinates while recording flake position | Movement of wafer or stage relative to each other after recording the global coordinates to set the origin (SW) and angle correction (SE) point. | After recording the flake position with reference to SW and SE, verify the origin position (SW) on the substrate. If it has been changed even a fraction of µm repeat the step 31(ii-iii) again. |
| 43 | Over development of lines | After EBL exposure, development of the lines pattern in developer for more than 30 s will lead to over development of spacers. | Get ready with the developer solution and IPA in separate beakers and set the time in the stopwatch. Take out the sample as soon as it reaches close to 30 s and dip the substrate in the IPA for 30 s. |
| 44 | Under and over-etching of lines | Under or over exposure to $O_2$ plasma for the etching of lines. | Etch the (EBL) exposed lines for right amount of time depending on the thickness of the spacer flake. |
| 45 | Difficulty in removing the PMMA | Continuous exposure of $O_2$ plasma on PMMA patterned spacer flake will result in crosslinking or making it difficult to remove later | Split the $O_2$ plasma exposure time into 20 s intervals with a 5-6 s wait in between each exposure cycle, to avoid cross-linking of PMMA. |
| 47 | Contamination of the capillary device | Stains or solvent residues left on the device due to the uncontrolled drying of device when transferred from acetone to IPA during the cleaning procedures. | When moving the device from acetone to IPA, avoid air dry which will result stains or solvent dry marks. If residue marks appear on the device, soak it in IPA for several minutes and transfer to another beaker of fresh IPA and repeat these steps if required. Dry with the compressed nitrogen gas. |
| 53 | Scratches on PMMA | Scratch marks on PMMA layer from tweezers, sample carrying boxes or while recording flake coordinates. | Avoid any unwanted scratches on PMMA surface. f happened than repeat PMMA coating steps and redo. |

| 58 | Folding of PMMA membranes during flake transfers. | Not picking up the film with appropriate tweezer. | Use flat tip tweezers. |
|---|---|---|---|
| 68 | Breaking of SiN$_x$ membrane during transfer | Transfer stage uses vacuum to hold substrates which may break the SiN$_x$ membrane if it is directly aligned to vacuum pinhole. | Stick Nitto tape to the backside of membrane before switching 'ON' the vacuum. |
| 108 | Hazard of chip breaking | Misaligned O-rings on the device chip while loading it in to the sample holder for He gas measurements. | Make sure the two O-rings are aligned on top of each other while loading the device chip into holder. |

**Timing**

Steps 1-22, fabrication of SiN$_x$ chips with a hole: 1-3 d

Steps 23-30, flake preparation and observation: 1-2 d

Steps 31-49, Spacer fabrication: 2-3 d

Steps 50-80, transfer of the flake: 2-3 d

Steps 81-107, Gold patch and device post-processing steps: 1-2 d

Steps 108-118, gas transport measurements: 1-2 d

Steps 109-125, water transport measurements: 3-5 d

Steps 126-144, *I-V* and streaming measurements: 1-3 d

**Anticipated Results**

**Device reliability:** The working conditions for each device are evaluated before their use. Once the device fabrication is complete, we test the devices by passing molecules through it via gas leak (Figs. 17 and 18) or electrochemical measurement techniques (Fig. 16) to check if the channels are open. If the channels were blocked or clogged, they do not allow gas or ions through them, hence the flux will be very low, close to that of a blank device.

Delamination and clogging can both compromise the channels' functioning. As discussed in a previous section, gold patches are used to clamp the channels on the membrane during the measurements, where osmotic or hydraulic stress, and voltage-induced delamination can occur on the nanochannels. Although minimized, delamination cannot be completely prevented and so we need to monitor the device's conductance from time-to-time. Fig. 16a compares the characteristic I-V curves for KCl (0.1 M at 25 °C) through a blocked, delaminated, and a good working channel device all with channels made of bilayer graphene spacer (~0.7 nm height). The primary difference is related to the conductance $G$ of the channels, which varies several orders of magnitude between these cases. The $G$ ranges can vary subject to experimental conditions, such as ion concentration, channel dimensions (height, width and length) along with number of channels. Scattered plot of the conductivity of the ionic conductivity $\sigma$ from several devices is shown in Fig. 16b; the yield of the working devices in the range of 30 % to 50 %. Because of the sub-nanometer dimensions of the channels, contamination from hydrocarbons either airborne or from the fabrication process, is a limiting factor leading to clogging of the channels. It is possible to revive the clogged channels (mostly but not in all cases) by annealing channel devices at 400 °C under $H_2$/Ar atmosphere. This gives the Å-channels a regenerative feature where the channel functionality is recovered. We can also extend their functioning by storing the devices in charcoal [95]. The He-leak tests were used to study the revival of clogged channels upon annealing (Fig. 17 a-b). Using hexane as a model molecule, contamination induced clogging of channels and recovery was found to be a function of the channel height. Graphene bilayer channels which are close to the size of the hexane did not make a full recovery whereas 5-layer thick channels were recovered by gas flushing and annealing [95].

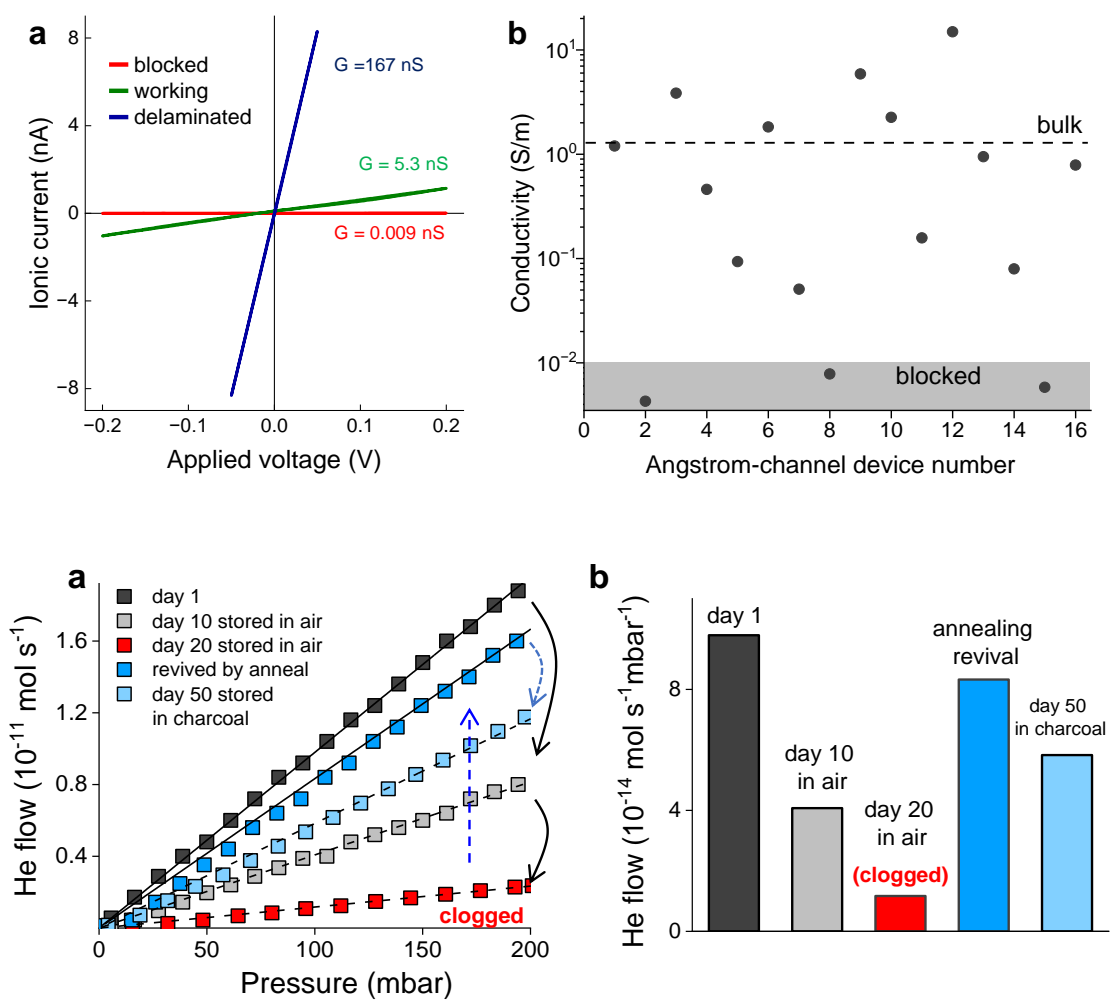

**Fig. 16| Anticipated results: evaluation of devices. a,** Typical I-V curves of 0.1M KCl solutions for bilayer channel (~ 0.7 nm thin) devices which are blocked (red curve), working (green curve), and delaminated (blue curve), respectively. Distinguishable differences up to several orders of magnitude in conductance ($G$) is seen for these devices. **b,** Scattered plot of the conductivity ($\sigma$) of several channel devices with various channel heights (from 0.7 nm to 2.8 nm) for 0.1 M KCl using Ag/AgCl electrodes. Data presented here illustrates the conductivity of 16 devices when measured just after fabrication. The $\sigma$ values are obtained from the $G$ (slope of the I-V curves) and the corresponding device's channel dimensions. Dotted line represents the bulk $\sigma$ of 0.1 M KCl at 25 °C (ref. CRC hand book [97]). Devices with σ values exceeding twice the bulk value were predominantly delaminated, whereas devices with

lower σ values within an order of magnitude were functional. The devices exhibiting σ values close to that measured with reference devices without channels (indicated by the grey coloured region) had blocked channels, while devices with σ values lower by more than an order of magnitude displayed only a few open channels.

**Gas flow measurement:** Various gas molecules including hydrocarbons can enter and flow through these 2D channels [95]. Particularly, He gas permeation through 2D channel devices is extensively investigated and Fig. 18 presents results for channels with thickness of five graphene layers [59]. It is expected that the He flow rate through the 2D channel device varies with the channel height. However, there is a peculiarly high-flow rate for the device with a spacer of few-layers of graphene as compared to thick spacer height channel device, which is attributed to specular reflection and ballistic transport [59] in thin channels. To further verify this result, He flow is measured through devices with the same spacer thickness (5-layer graphene), but with confining walls made from different 2D materials (graphite, hBN and $MoS_2$). Compared to the gas flow rate value calculated from the Knudsen equation, enhanced permeation is observed for devices made with graphene and hBN walls, while $MoS_2$ gave a flow rate closer to the Knudsen value. As both hBN and graphene exhibit atomically flat surfaces while $MoS_2$ does not, this hints at the specular reflection of gas molecules. To further confirm this, several channel devices with graphite and $MoS_2$ walls were made with varying channel lengths. The gas flow in $MoS_2$ channels is an inverse function of channel length as expected from the Knudsen equation, whereas graphite channels show length-independent flow. This confirms that specular reflection on smooth walls of graphite and hBN can cause enhanced flow and thus giving a high leak rate.

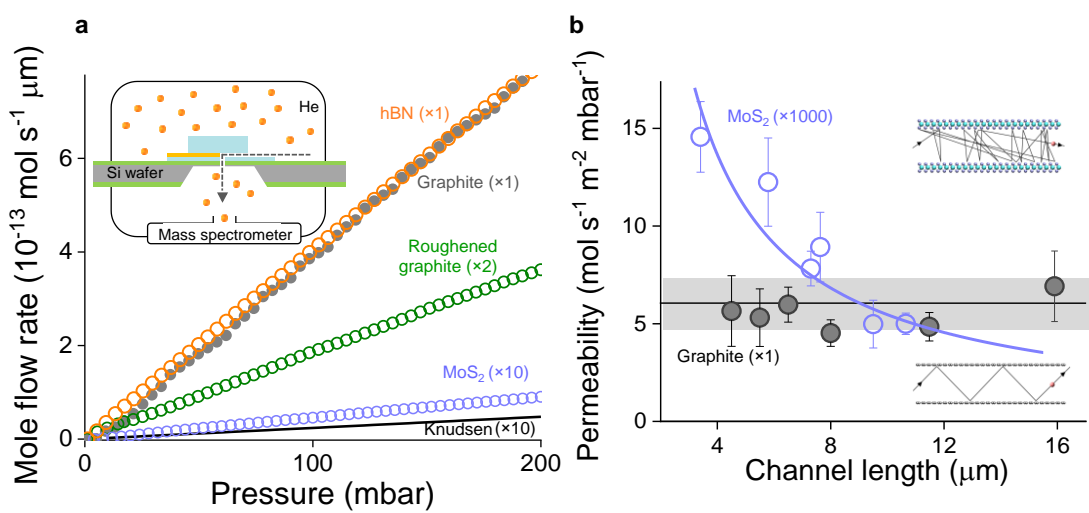

**Fig. 17| Anticipated results: Effect of clogging and regeneration of devices. a,** Storage conditions of a 5-layer graphene spacer (height, h ~ 1.7 nm) channels device (with graphite top and bottom walls) and Helium flow rate. There are ~200 channels in this device with an average channel length of ~5 μm. **b,** Bar graph referring to the data in panel (**a**) shows revival of the devices upon annealing. By storing the channels in ambient conditions, a significant reduction in flux is observed within a few days, which is due to the channel clogging. However, the flow can be recovered by annealing at 400 °C. Device stored in charcoal over ~50 days shows minimal flow reduction [95].

**Water flow measurement:** The channels with heights as small as mono layer graphene spacer to few tens of nanometers allow water permeation [99] and results from devices with graphite walls (top and bottom) are presented in Fig. 19 [57]. The water permeation rate for the monolayer channels is much less (one order or lower) than the flow through bilayer Å-channels probably due to highly ordered nature of confined water [60]. Given the small cross-sectional area of the Å-channels, to be able to detect such measurable flows, there might be an evaporating extended meniscus at the exit of the channels. In a nanoscale opening, the extended meniscus is likely to be composed of a thin layer of water driven by high spreading pressure over micrometre distances (Fig.4b). There is a high-water permeation rate (weight loss rate on Fig. 19a) for the channels with heights 1.3 nm and 1.7 nm graphene compared to thicker (10 nm) channels, which is attributed to the combined effect of capillary pressure and disjoining pressure built rapidly inside the channels [57]. We can see decreasing water permeation for the channel heights above 1.7 nm due to a less pronounced disjoining pressure. Once the channel height exceeds 3 nm, the flow rate increases with almost linear dependence on the channel height, with the flows close to that estimated from the slip-corrected Hagen-Poiseuille equation.

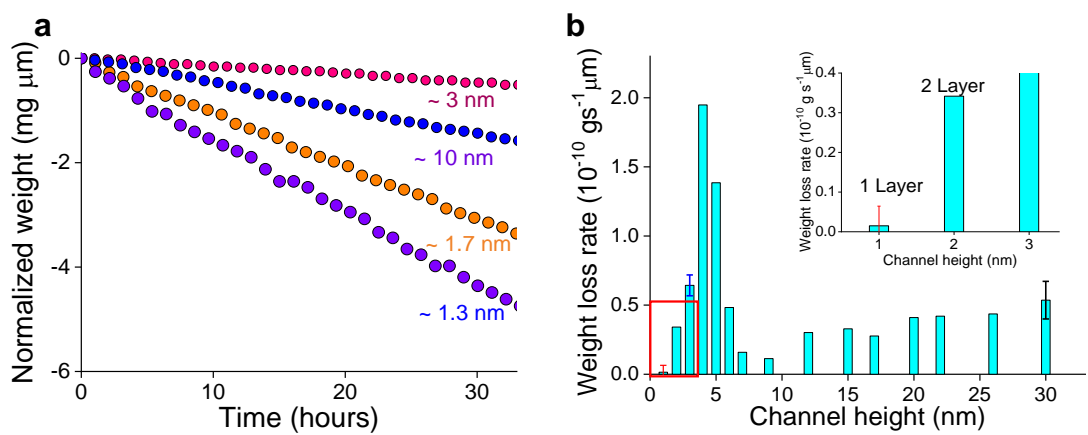

**Fig. 18| Anticipated results: gas flow measurement through the 2D channels device**. **a,** He flow through 2D channel device with 5-layer graphene spacer (h ∼ 1.7 nm) but with walls made of graphite, hBN and $MoS_2$ compared to the expected Knudsen flow rate (black line). For legibility, mole flow rates are multiplied by the factors shown in the colour-coded brackets. Ballistic transport is observed for devices with graphite and hBN walls. **b,** Helium flow through channels comprising 4-layer graphene spacer, graphite or $MoS_2$ walls shown as a function of channel length $L$. The flow rate divided by P×w×h gives the permeability. Due to the diffusive transport (top inset) of gas molecules on $MoS_2$ walls, the permeability shows the $1/L$ dependence; the purple curve is a fitting to the $1/L$ data. In graphene channels, there is no length dependence, which indicates frictionless transport due to specular scattering (bottom inset). $MoS_2$ and graphite datasets are shown with purple and black symbols, respectively, indicating standard error in measurements using at least two devices for each length. Adapted from reference [59].

**Ion transport measurements**: We now discuss the key features of the ion flow under atomic scale confinement with the application electrical or mechanical stimuli.

**Steric exclusion of ions:** The confinement along the height of the channels can lead to exclusion of ions with hydrated diameter larger than the channels. Ion conductance was tested for different ions by passing a series of chloride salts through a device with Å-channels of height equivalent to 2 layers of graphene fabricated with the presented protocol in ref [58]. Surprisingly the ions with a hydrated diameter larger than the channel height can still permeate albeit with reduced mobility, which suggests partial dehydration of ions. Under Å-scale confinement, the largest cations studied exhibit a 10-fold decrease in mobility [58]. When the channel height was reduced further down to one graphene layer, only protons could permeate through the channel, as shown in Fig. 20b. All the monovalent and multivalent cations were excluded from Å-channels with their measured ionic conductance below the detection of limit. In the case of HCl, even though the protons permeate through monolayer Å-channels, their mobility was still much smaller that in the bulk [60].

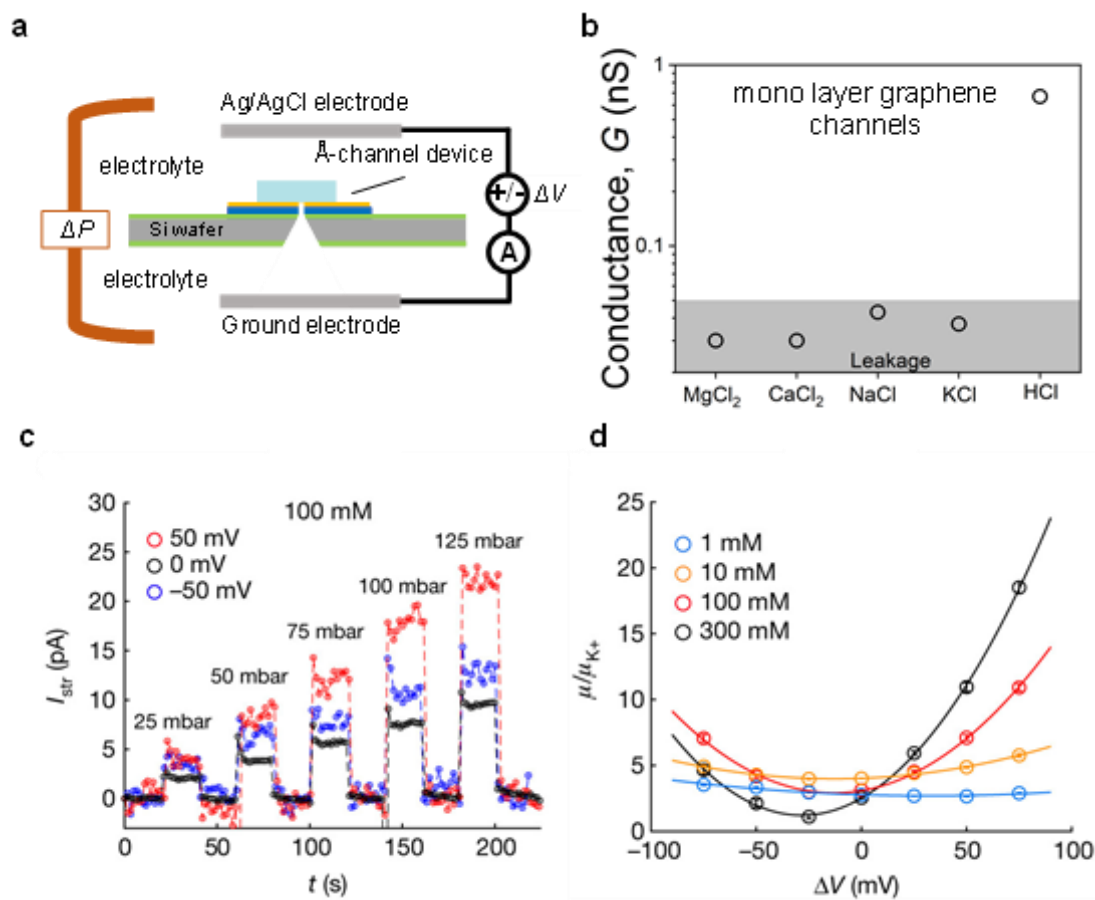

**Fig. 20| Anticipated results: ion and pressure driven transport measurements in Å-channels. a,** Schematics of the experimental setup for measuring voltage-driven and pressure-driven currents. A known pressure ΔP is applied to KCl electrolyte from one end and the resulting pressure-induced current is measured with Ag/AgCl electrodes. **b**, Ionic conductivity measured for the monolayer thin channels ($N = 100$; $L = 10$ μm; hBN/graphene/hBN) using 1M solutions. Grey areas represent current detection limit measured using reference devices and blank Si nitride membranes. **c**, The streaming current $I_{str}$ as a function of time for graphite channels; c= 100 mM KCl; $L = 5.7 \pm 0.1$ μm. Current overshoots once the pressure is applied, and only the steady-state regime was considered in this study. The pressure applied for 20 s intervals is gradually increased to 125 mbar in 25 mbar steps. **d,** Voltage gating study for various concentrations of KCl using graphite devices - streaming mobility (normalized by the $K^+$ electrophoretic mobility) as a function of ΔV. Curves in **d** are the quadratic fits [61]. The graphs in **c** and **d** are adapted from reference [61].

**Streaming current:** The ionic current generated with pressure-driven flow of electrolyte through the channels is known as streaming current [99]. Data presented in Fig 20c shows the pressure-induced streaming current in atomic scale channels. A positive streaming current indicates that the flow carries a pressure-dependent net positive charge, where anion mobility is suppressed under the strong confinement. The streaming currents can be further modified with an additional applied voltage of only few tens of millivolts (gating effect) [61]. The magnitude of the streaming current is thus a function of the applied pressure, voltage and the ion concentration. The streaming mobilities measured in graphite channels and normalized by K$^+$ electrophoretic mobilities are shown in Fig. 20d as a function of ΔV and ionic concentration. The ionic streaming currents and the gating effects are sensitive to the surface of the channels – while graphite walls produce a gating effect of up to 20 times increase in mobility and quadratic dependence on the gating voltage, hBN channels show a linear dependence and only few times increase of mobility with gate voltage [61].

Applications of these Å-scale channels fabricated using our method are not limited to the type of studies we presented here. The method may contribute to an emerging field in condensed matter physics and chemistry where a series of phenomena yet to be explored at angstrom-scale are shaped by such 2D empty spaces. For instance, our channels have been used to study DNA translocation [63] and long-term memory effects as ionic memristors at sub-nanometer scale [100]. In both studies, the complexity of these phenomena could be studied by improving our understanding of atomic scale fluidic systems, *via* the detection of features that could not be observed or considered previously. In addition, for future studies, our Å-channels may be used as templates for fabricating other types of systems, such as solid-state metallic or inorganic materials, making it possible to study their physical properties at unprecedented scales.

## Acknowledgements

AK acknowledges Royal Society research grants (RGS\R2\202036 and IES\R3\203066), EPSRC new horizons grant (EP/V048112/1). B.R. and A.K. acknowledge EPSRC strategic equipment grant (EP/W006502/1). B.R. acknowledges funding from the Royal Society University Research Fellowship (URF\R1\180127, RF\ERE\210016), Philip Leverhulme Prize PLP-2021-262, EPSRC New Horizons grant EP/X019225/1, funding from the European Union's H2020 Framework Programme/European Research Council Starting Grant (852674 – AngstroCAP). R.Q acknowledges CSC scholarship. All authors are thankful to Mark Sellers and Dominic Mccullagh for their technical support for custom design and manufacturing of electrochemical cells and gas transport holders.

## Author contributions

A.K., A.B., Y.Y., R.S., R.Q., S.A.D. and M.R. carried out the fabrication of Å-channel devices. A.K., R.B., A.B., R.S. and M.V.S.M. carried out the device characterization. S.G. performed the ion conductance measurements and their analysis. A.B., M.V.S.M., Y.Y., B.R. and A.K. wrote the manuscript with inputs from M.R., R.S. and S.G. All the authors contributed to discussions. A.K. and B.R. provided supervision for the protocol.